# Legacy Survey of Space and Time cadence strategy evaluations for active galactic nucleus time-series data in Wide-Fast-Deep field

Xinyue Sheng ![ORCID],[1][*] Nicholas Ross ![ORCID],[2] and Matt Nicholl ![ORCID][1]

[1]*Institute of Gravitational Wave Astronomy and School of Physics and Astronomy, University of Birmingham, Birmingham B15 2TT, UK*
[2]*Advanced Research Division, Niparo, 41 Dundas Street, Edinburgh EH3 6QQ, UK*

## ABSTRACT

Machine learning is a promising tool to reconstruct time-series phenomena, such as variability of active galactic nuclei (AGNs), from sparsely sampled data. Here, we use three Continuous Autoregressive Moving Average (CARMA) representations of AGN variability – the Damped Random Walk (DRW) and (over/under)Damped Harmonic Oscillator – to simulate 10-yr AGN light curves as they would appear in the upcoming Vera Rubin Observatory Legacy Survey of Space and Time (LSST), and provide a public tool to generate these for any survey cadence. We investigate the impact on AGN science of five proposed cadence strategies for LSST's primary Wide-Fast-Deep (WFD) survey. We apply for the first time in astronomy a novel Stochastic Recurrent Neural Network (SRNN) algorithm to reconstruct input light curves from the simulated LSST data, and provide a metric to evaluate how well SRNN can help recover the underlying CARMA parameters. We find that the light-curve reconstruction is most sensitive to the duration of gaps between observing season, and that of the proposed cadences, those that change the balance between filters, or avoid having long gaps in the $g$ band perform better. Overall, SRNN is a promising means to reconstruct densely sampled AGN light curves and recover the long-term structure function of the DRW process (SF$_\infty$) reasonably well. However, we find that for all cadences, CARMA/SRNN models struggle to recover the decorrelation time-scale ($\tau$) due to the long gaps in survey observations. This may indicate a major limitation in using LSST WFD data for AGN variability science.

## 1 INTRODUCTION

The stochastic variability of quasars has been recognized and studied since the discovery of the active galactic nucleus (AGN) phenomenon[1] (Greenstein 1963; Hazard, Mackey & Shimmins 1963; Matthews & Sandage 1963; Oke 1963; Schmidt 1963; Press 1978). Observations of large numbers of quasars from wide-field surveys show changes in the ultraviolet (UV)/optical emission (e.g. Vanden Berk et al. 2004; Wilhite et al. 2005; Sesar et al. 2007) and at higher energies (e.g. Tarnopolski et al. 2020) over the course of weeks to decades. It is now understood there are several physical mechanisms that underlie the observed quasar variability, with the variable continuum UV-optical emission driven by thermal fluctuations in the active accretion disc (e.g. Shakura & Sunyaev 1973; Czerny et al. 1999; Peterson & Horne 2004; Kelly, Bechtold & Siemiginowska 2009; Kelly et al. 2014). Understanding the details of the AGN accretion disc is a key area of study in contemporary astrophysics (e.g. Dexter & Agol 2011; Cai et al. 2016; Dexter & Begelman 2018; Kubota & Done 2018; Jiang et al. 2019; Jiang & Blaes 2020).

A time Continuous-Autoregressive (C-AR) process has been proposed for describing the observed UV/optical AGN variability

(e.g. Brockwell & Davis 2002; Kelly et al. 2009; Kozłowski et al. 2010). This was later extended to a more flexible and scalable model – the Continuous Autoregressive Moving Average (CARMA) model (Kelly et al. 2014; Feigelson, Babu & Caceres 2018; Moreno et al. 2019). CARMA models are not physical models, but rather a statistical description that characterizes a time-series stochastic process. CARMA models are notated as CARMA($p$, $q$) where $p$ gives the order of the Autoregressive (AR) process and $q$ gives the description of the Moving Average (MA) process. The AR response can be thought of as the forecasting part, while the MA model gives the input impulse(s). The Power Spectral Density (PSD),[2] the Autocorrelation Function (ACF),[3] and the Structure Function (SF)[4] can all be calculated for CARMA models. Moreno et al. (2019) present a detailed CARMA handbook for optical AGN variability, discuss CARMA models and their associated statistical descriptions in full detail, and illustrate the bridge between discrete ARMA and time-continuous CARMA for fitting the irregular sampling of light curves.

---

[1]We use the term quasar and AGN interchangeably in the manuscript, noting that quasars are the bolometrically luminous subset of AGN.

[2]The power spectrum of the signal, describing the distribution of power across frequencies, i.e. the Fourier transform.
[3]The correlation between steps in a time series.
[4]The average change as a function of time interval.

The simplest CARMA model, CARMA(1,0) is the well-known Damped Random Walk (DRW). The DRW, and its ACF and SF can be expressed as

$$d^1 x + \alpha_1 x(t) = \beta_0 dW(t),\tag{1}$$

$$ACF(\Delta t) = e^{-\frac{|\Delta t|}{\tau}},\tag{2}$$

$$SF(\Delta t) = SF_\infty (1 - e^{-|\Delta t|/\tau})^{1/2}, \; SF_\infty = \sqrt{2}\sigma,\tag{3}$$

where $\alpha_1$ is the C-AR coefficient and $\beta_0$ is the coefficient of the random perturbations. In the case of AGN, $x$ corresponds to the flux or magnitude. $W(t)$ is a Wiener process, and $dW(t)$ means a white noise process with $\mu = 0$ and variance $= 1$ (Kelly et al. 2014). $\Delta t$ is the difference between two MJDs. There are two parameters that can be obtained from DRW to capture the statistics of AGN variability: $\tau$, the characteristic damping (or signal decorrelation) time-scale, and $SF_\infty$, the long-term variability amplitude.

The DRW model is a good description of long-term quasar variability and as such is often applied to data from large-area sky surveys. MacLeod et al. (2010) confirmed that a DRW model fits well for ~9000 Sloan Digital Sky Survey (SDSS) Stripe 82 quasar light curves, and analysed correlations between the observed variability parameters and AGN physical parameters including black hole mass, redshift, luminosity, and rest-frame wavelength of emission. Suberlak, Ivezić & MacLeod (2021) built on this work by adding Pan-STARRS1 photometry to the SDSS Stripe 82 data, generating light curves up to 15 yr in length. They found that the variability amplitude is a stronger function of the black hole mass, and that it (and $\tau$) has a weaker dependence on quasar luminosity than initially found in MacLeod et al. (2010).

Kozłowski (2016b, 2017) investigated systematic biases (photometric noise, etc.) in SF measurements. They applied Monte Carlo simulations to AGN light curves, and showed that accurate estimation of DRW parameters requires the observation sampling to be at least 10 rest-frame decorrelation time-scales. Kozłowski (2017) note this is because observations shorter than ~10$\tau$ are insufficient to fully sample the PSD. Thus, due to the limited observation lengths of astronomical surveys, the estimations of DRW parameters may fall into this unconstrained region resulting in biases.

Kozłowski (2016a) reported that DRW modelling can also work well for non-DRW processes, and should not be regarded as a proxy for the physical process underlying the variable emission. On shorter time-scales, other models may be more appropriate. Kasliwal, Vogeley & Richards (2015) studied a number of Kepler AGN light curves (Howell et al. 2014). Compared with ground-based surveys, Kepler has higher photometric precision and denser observation cadences, but shorter survey length. Kasliwal et al. (2015) found that AGN with Kepler light-curve information had log-PSD slopes steeper than that of DRW, suggesting that the variability may be better captured by another process other than AR(1).

Combining SDSS and data from Kepler's second mission, K2, Kasliwal, Vogeley & Richards (2017) discuss DRW and a higher order CARMA(2,1) model, the Damped Harmonic Oscillator (DHO), and indicate that an overdamped DHO[5] may be a better description of the AGN Zw 229-15 (see figs 1 and 3 from Kasliwal et al. 2017). The PSD time-scale features for Zw 229-15 are also reported by Edelson et al. (2014) and Williams & Carini (2015). Kovačević, Popović & Ilić (2020) present a method to model AGN variability using a representation of the DHO model with Gaussian processes

(GPs), and successfully detect variability due to continuum emission and (broad) line emission.

Thus, DRWs and DHOs are useful descriptions of quasar light curves. The key goal of CARMA models now is to accurately measure the model parameters from observed quasars, and use this information to study the underlying physics. We summarize the working equations for the DRW=CARMA(1,0) and DHO=CARMA(2,1) models in Table A1. Table A1 presents the differential equations, input parameters, impulse response (IR, also called Green's function), SF, ACF, Autocovariance Function (ACVF), and PSD for DRW and DHO processes. Table A2 explains all acronyms and notation. Table 1 shows all acronyms used in this paper.

Our ability to study AGN variability will soon be transformed by the Vera Rubin Observatory (VRO), conducting the LSST. The *VRO* telescope, located on Cerro Pachon in Chile, has an 8.4-m (6.5-m effective) primary mirror, with a 9.6-deg$^2$ field of view, a 3.2-gigapixel camera, and six filters (*ugrizy*) covering the wavelength range 320–1050 nm.

LSST will repeatedly observe millions of objects, with $\geq 825$ visits for any given point in the survey footprint and a single-visit depth $\approx 24.5$ mag in the $r$ band. LSST will cover the whole Southern sky, and part of the Northern sky, as part of the Wide-Fast-Deep (WFD) survey. There are also five Deep Drilling Fields (DDFs; LSST Science Collaboration et al. 2009). The WFD will take about 90 per cent observing time, with >10 million quasars projected to be identified, though the cadence strategies will have an effect on the efficiency of quasar identification (Ivezic 2016). This has motivated recent white papers from the LSST AGN Science Collaboration for estimating the influence of cadence strategies on AGN astrophysics studies.[6] Kovacevic et al. ([2021b](#)) have provided statistical proxies to measure the LSST cadence effects on AGN variability observations and Kovacevic et al. ([2021a](#)) provide two metrics: based on AGN time lag and periodicity, and on the SF. . Such models simulate AGN light curves ahead of the start of LSST survey operations. Analysing recovery of parameters (including e.g. characteristic time-scales, PSDs, and IRs) from the simulated data under different cadences can tell us the potential systematic bias.

**Table 1.** Acronyms used in this paper.

| | |
|---|---|
| **Variability models:** | |
| ACF | Autocorrelation Function |
| ACVF | Autocovariance Function |
| CARMA | Continuous Autoregressive Moving Average |
| DHO | Damping Harmonic Oscillator |
| DRW | Damped Random Walk |
| PSD | Power Spectral Density |
| SF | Structure Function |
| **LSST related:** | |
| DDF | Deep Drilling Fields |
| LSST | Legacy Survey of Space and Time |
| OpSim | Operations Simulator |
| VRO | Vera Rubin Observatory |
| WFD | Wide-Fast-Deep (survey) |
| **Data science terms:** | |
| AE | Auto-Encoder |
| ELBO | Evidence Lower Bound |
| GP | Gaussian processes |
| GPR | Gaussian process regression |
| ML | Machine Learning |
| RNN | Recurrent Neural Network |
| RAE | Recurrent Auto-Encoder |
| SRNN | Stochastic Recurrent Neural Network |
| SSM | State Space Model |
| VAE | Variational Auto-Encoder |
| VRAE | Variational Recurrent Auto-Encoder |
| LSTM | Long Short-term Memory Unit |
| GRU | Gated Recurrent Unit |

[5]Whose IR moves slowly toward equilibrium.

With the development of large sky surveys such as SDSS and LSST, astronomy has become a data-intensive science. Consequently, there have been recent attempts to use ML techniques to address the data challenges set by LSST, especially for classification, forecasting, and parameter estimations. For example, the Photometric LSST Astronomical Time Series Classification Challenge included 175 500 simulated AGNs (among other transients) to test classification algorithms (e.g. Boone [2019](#); Kessler et al. [2019](#); Hložek et al. [2020](#)). Relevant to the AGN study (Jankov et al. [2022](#)) based on LSST cadence strategies, in this paper, we aim to quantify the influence of different LSST cadence strategies on AGN time-series data, in order to see if contemporary ML algorithms can effectively recover CARMA model parameters.

We implement an SRNN to model quasar light curves and recover the DRW and DHO model parameters by GPR. RNN are a popular class of ML connectionist models for sequential modelling and have been used previously in astrophysics applications (e.g. Charnock & Moss [2017](#); Hinners, Tat & Thorp [2018](#); Naul et al. [2018](#); Muthukrishna et al. [2019](#); Becker et al. [2020](#); Escamilla-Rivera, Carvajal Quintero & Capozziello [2020](#); Möller & de Boissière [2020](#); Burhanudin et al. [2021](#); Lin & Wu [2021](#)). However, as noted in Yin & Barucca ([2021](#)), one limitation of RNNs is that the hidden state transition function is entirely deterministic, which can limit the RNNs ability to model processes with high variability. Thus (and as far as we can tell, for the first time in astrophysics research), we implement the SRNN, in order to recreate quasar light curves. The SRNN differs from the traditional RNNs in that the RNN hidden cells invoke a probabilistic (often Gaussian) distribution to generate a level of stochasticity that improves longer term temporal forecasting. As such, the SRNN can be somewhat thought of as a combination of an RNN and ideas from a VAE.

We have two main goals:

(i) To test how well the SRNN can recover and predict observations when dense or uniformly seasonal light curves are set as inputs.

(ii) To predict AGN behaviour during gaps between seasons in 10-yr LSST-simulated light curves, and see how SRNN could help recover DRW and DHO parameters.

This paper is organized as follows. In Section [2](#), we describe our observational data and the methods (including DRW and DHO) we use to generate sample quasar light curves. Section [3](#) presents the details on the LSST observing strategy and associated cadences. In Section [4](#), we describe the ML algorithms we use to evaluate the model quasar light curves. We report our key results in Section [5](#) and discuss these results in the context of quasar studies and LSST in Section [6](#). Section [7](#) presents our conclusions. In Appendix [A](#), we write down the fundamental parameters and equations for the CARMA models, and in Appendix [B](#), we detail the implementation of the SRNN. We report all magnitudes on the AB zero-point system

(Oke & Gunn [1983](#); Fukugita et al. [1996](#)) unless otherwise stated. All logarithms are to the base 10.

## 2 QUASAR DATA AND MODEL LIGHT CURVES

In this section, we give the details of our quasar data, including both observed quasars from previous sky surveys, used to provide an input set of representative statistical parameters, and the model light curves we generate. For the observed quasar data we will focus on the well-studied SDSS Stripe 82 field. The key analysis codes of this section are available via a github repository.

### 2.1 Observed quasars

#### 2.1.1 SDSS Stripe 82

The Stripe 82 field (hereafter [S82](#); Annis et al. [2014](#)) is a ∼300-deg² region of the SDSS across 22h 24m < RA < 04h 08m, |Dec| < 1.27 deg, and has been observed ∼60 times on average to search for transient and variable objects (Abazajian et al. [2009](#)). These multi-epoch data have time-scales ranging from 3 h to 8 yr and provide well-sampled five-band light curves for an unprecedented number of quasars. Examples of quasar variability studies based on [S82](#) photometry include Sesar et al. ([2007](#)), Schmidt et al. ([2010](#)), Ai et al. ([2010](#)), MacLeod et al. ([2010](#)), MacLeod ([2012](#)), Meusinger, Hinze & de Hoon ([2011](#)), Butler & Bloom ([2011](#)), Kozłowski ([2016b](#)), and Suberlak et al. ([2021](#)).

For our study we will concentrate on the ∼9000 spectroscopically confirmed quasars from SDSS [S82](#) reported in MacLeod et al. ([2010](#)).[7] MacLeod et al. ([2010](#)) model the time variability using the DRW and measure the characteristic time-scale ($\tau$) and an asymptotic rms variability on long time-scales (SF$_\infty$). Also reported for this data set is the binary value edge (if the observation is close to the field edge) and a set of probabilities: $P$like (log likelihood of a DRW solution), $P$noise (log likelihood of a noise solution), and $P$inf (log likelihood of $\tau \rightarrow \infty$).

In order to have a sample of objects that have well-fitted DRW parameters, we selected those with edge = 0, $P$like-$P$noise>2, and $P$like-$P$inf>0.05. We also remove objects with $\tau < 0$ or $\tau > 10^5$ d to allow convergence in $\tau$. With these selections, the number of quasars in the sample is 7384. We plot the SF$_\infty$ and $\tau$ distribution of these quasars in Fig. [1](#).

### 2.2 Model quasars

Our aim is to analyse the influence of survey cadence on quasar modelling in LSST. We note again that although CARMA models are not physical models, they are appropriate approximations of quasar light-curve properties. As such, we simulate the light curves using a DRW and DHO implementation. We first generate 10-yr light curves (consistent with LSST survey length) with dense, daily observations, which can later be sampled at different realistic cadences. The steps to generate light curves are described in the following sections.

#### 2.2.1 DRW-simulated light curves

For the DRW model, two input parameter choices are required: SF$_\infty$ and $\tau$, in addition to the redshift ($z$). To generate our model light curves, we sample the SF$_\infty$-$\tau$ parameter space from the [S82](#) data set

---

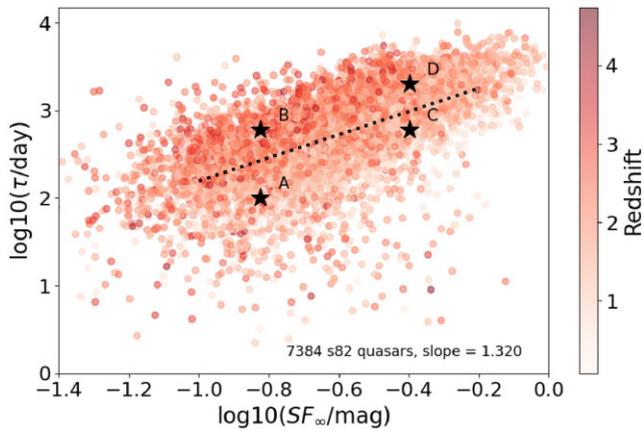

**Figure 1.** The $SF_\infty$ and $\tau$ distribution of 7384 quasars in $g$ band selected from the SDSS S82 field, with spectroscopic redshift given by point colour. The dashed line is the regression line, and the four star markers represent four DRW parameter pairs, the associated light curves of which are shown in Fig. 2.

(MacLeod et al. 2010) shown in Fig. 1. We take the values reported at a given $SF_\infty$–$\tau$ coordinate and add a 'scatter' in the range $-0.05 < \iota < 0.05$ (as determined by a random number generator) to each ordinate separately. We generate 30 000 DRW light-curves parameter pairings in this manner.

The four black stars in Fig. 1 are four DRW pairs, and we show their associated light curves, SFs and PSDs in Figs 2 and 3. Given a fixed value for $SF_\infty$, and a set rest-frame observation length, longer characteristic time-scales ($\tau$) will lead to less fluctuation. When $\tau$ is short compared with the rest-frame observing duration, the variance of the light curve will tend towards $\sigma^2$ and the estimated parameters will approach the underlying values.

### 2.2.2 DHO-simulated light curves

The literature is not so comprehensive for DHO modelling of UV/optically bright AGN, though there are papers that report the PSD slopes of some quasars that are inconsistent with DRW (e.g. Kasliwal et al. 2015, 2017; Moreno et al. 2019). Consequently, these illustrate that a more complex CARMA process, DHO, might be more suitable for describing some quasars with (quasi-)periodicity features, such as weak oscillations, though such research is purely based on the statistical analysis of their variability rather than any assumptions of deterministic/periodic physical processes. As such, we build a DHO set including both 'overdamped' and 'underdamped' cases.

Five parameters are required as inputs: $\beta_0$, $\beta_1$, $\xi$, $\tau_{decay}$, and $\tau_{QPO}$ [the time-scale of quasi-periodic oscillations (QPOs)] in the 'underdamped' case or $\tau_{rise}$ (the time-scale to brighten in response to an impulse) in the 'overdamped' case (see Table A2 for details). To build the light curves, the redshift and $SF_\infty$ for each light curve are randomly sampled from the S82 distribution. $\tau_{decay}$ is set to vary from 60 to 200 d. MA coefficients $\beta_0$ and $\beta_1$ are set with constant and small values to ensure that dependent parameters remain in reasonable ranges (see Appendix A1). Additionally, we set $2 < \xi < 5$ for the overdamped case, $0 < \xi < 1$ for the underdamped case. In this way, we can calculate the corresponding C-AR coefficients $\alpha_1$ and $\alpha_2$ using the recipes in Table 2.

Fig. 4 shows four DHO light curves, including two underdamped and two overdamped, with their IR functions. For the underdamped DHO, the IR oscillates and gradually returns to a steady state, whereas in the overdamped case, it gradually moves to its steady state without oscillation. Analogous to Figs 2 and 3 for the DRW model, Fig. 5 depicts the PSD and SF for the underdamped and overdamped DHO.

### 2.2.3 GPRs and Eztao

A GP is a generalization of the Gaussian probability distribution and a GPR model provides uncertainty estimations together with prediction values. GP[8] and GPR are discussed extensively elsewhere (e.g. Rasmussen & Williams 2006).

CARMA model can be well expressed by a GP model which consists of a mean function and a covariance matrix (also called kernel). For example, the kernel for the simplest CARMA process (DRW) can be written as

$$k(\Delta t) = \sigma^2 e^{-\Delta t/\tau}. \tag{4}$$

We generate the DRW and DHO light curves using the Eztao Python package (Yu & Richards 2022). Eztao is a Python toolkit for conducting time-series analysis using CARMA processes. Building on work by Rybicki & Press (1995) and in particular Foreman-Mackey et al. (2017), EzTao uses celerite (a fast GPR library) to compute the likelihood of a set of proposed CARMA parameters given the input time series.

Here, we use EzTao to model and produce GPRs that give uncertainty estimations together with predictions for DRW and DHO light curves.

### 2.2.4 Colour–redshift correlation and $SF_\infty$

The observed colour of a quasar changes with redshift (e.g. Richards et al. 2001). To quantify this, we select 151 362 quasars included in the SDSS DR16 Quasar catalogue (Lyke et al. 2020) that are also in the UKIDSS (Lawrence et al. 2007) footprint. We convert from UKIDSS/WFCAM[9] Vega $Y$-band to LSST AB $y$-band,[10] and calculate the colours $(u - g)$, $(g - r)$, $(r - i)$, $(i - z)$, and $(z - y)$ from redshift $z = 0.00 - 5.00$, in redshift bins of $\Delta z = 0.05$. We initially generate mean magnitudes in the $g$ band, and normalize the other bands using these colour relations.

Examining the $SF_\infty$ values for 9258 S82 quasars from MacLeod et al. (2010),[11] we find that $SF_\infty$ is larger in bluer bands. As SDSS does not provide a $y$ filter, we calculated the $SF_\infty$ and $\tau$ ratios for the S82 quasars in six bands (shown in Table 3), and fix the ratio of $y$ band $SF_\infty(y)/SF_\infty(u) = 0.61$, and $\tau(y)/\tau(u) = 1.26$.

### 2.2.5 LSST photometry and photometric error

Here, we note the expected photometric performance of LSST, and tie this to our synthetic model quasar light curves. Detailed LSST performance metrics are given in Ivezić et al. (2019, section 3.2.1), and the expected photometric error in magnitudes for a single visit can be written as

$$\sigma^2_{LSST} = \sigma^2_{sys} + \sigma^2_{rand} \tag{5}$$

$$\sigma^2_{rand} = (0.04 - \gamma)x + \gamma x^2 (mag^2), \tag{6}$$

---

[9] The wide-field camera on the *United Kingdom Infrared Telescope*.
[10] $mag_{AB} = -2.5\log(F_\nu) - 48.6$
[11] http://faculty.washington.edu/ivezic/macleod/qso_dr7/Southern.html

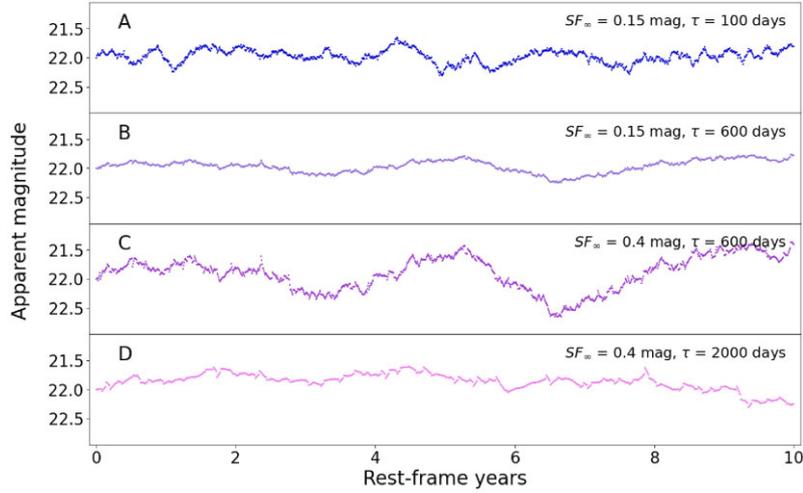

**Figure 2.** Four light curves simulated from the DRW parameters in the SDSS S82. The mean magnitude is 22.0 mag.

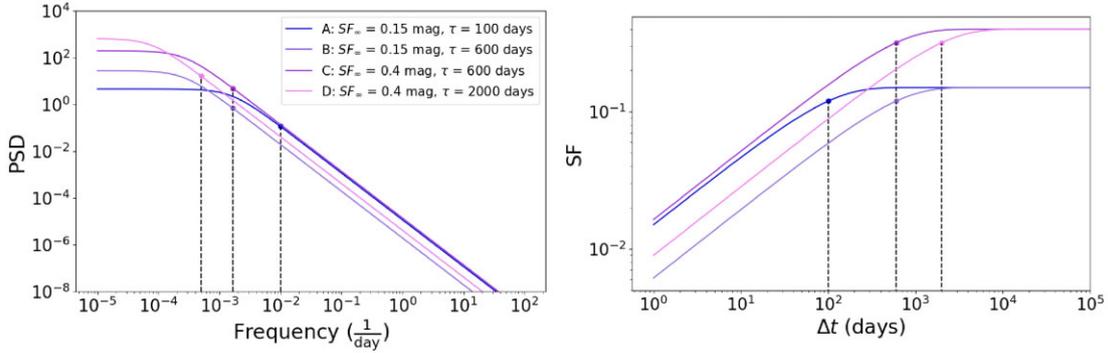

**Figure 3.** Left-hand panel: The PSD for the four example DRW light curves as given in Fig. 2. The PSD is determined by the driving force [$\beta_0 \, dW(t)$ in DRW equation]. For DRW, there is only one AR coefficient $\beta_0$, indicating that slope of PSD will be flat with $f^{-2}$ after the break point. Right-hand panel: The SF describes the trend of standard variance of magnitude differences with $\Delta t$. The vertical dash lines correspond to $\tau^{-1}$ and $\tau$ in each graph, respectively.

**Table 2.** DHO parameters for underdamped and overdamped cases. These values restrain the range of the real parameters, but they are not applied with fixed increments to avoid combinations that lead to unrealistic derived statistical parameters (calculated using Table A1) – such simulated cases are removed.

| Parameter | Values | Values |
|---|---|---|
| | Underdamped | Overdamped |
| $\tau_{decay}$ | 60~200 | 60~200 |
| $\tau_{QPO}$ | 7.8~380 | – |
| $\tau_{rise}$ | – | 5.6~38.8 |
| $\xi$ | 0~1 | 2~5 |
| $\beta_0$ | 0.0022 | 0.003 672 |
| $\beta_1$ | 0.000 25 | 0.0257 |
| $N_{tot}$ | 1000 | 1000 |

where $\sigma_{sys}$ and $\sigma_{rand}$ are the systematic and random photometric error, respectively, $x \equiv 10^{0.4(m-m_5)}$ and $m_5$ is the typical $5\sigma$ depth of point source at zenith for each visit. Given the fact that the calibration system and procedure are set to maintain $\sigma_{sys} < 0.005$ mag, we assume $\sigma_{sys} = 0.004$ mag. $\gamma$ is a band-dependent parameter. Following Table 2 in Ivezić et al. (2019), we set $\gamma_u = 0.038$ and $\gamma_{g,r,i,z,y} = 0.039$. LSST Operations Simulator (OpSim) provides the $m_5$ for each visit in a proposed cadence strategy. In this way, for one original simulated photometric value, a photometric error is randomly selected from a Gaussian distribution, with mean equal to 0 and variance equal to $\sigma_{LSST}^2$.

### 2.3 Model light-curve data products

Following the steps above, we generate a data set containing light curves generated from the DRW and DHO models. We simulate 200 objects from each of the DRW, DHO-overdamped and DHO-underdamped models, and for each object, light curves in six bands are generated, with one observation per day. The mean magnitude, $SF_\infty$, and $\tau$ follow the restrictions detailed in Section 2.2. These light curves will be used in Section 4 for the SRNN analysis, and are also made available to the community.[12]

## 3 SIMULATED AGN WITH LSST CADENCES

In this section, we present the light curves of the simulated quasars and how they will be observed using realistic LSST observing cadences. The survey strategy and cadence choices for LSST are described in detail by Jones et al. (2021; LSST Project Science Team Note 051). The LSST survey strategy is designed to fulfil the core science goals (which can be found at Science Requirements

[12]https://github.com/XinyueSheng2019/LSST_AGN_SRNN_Paper

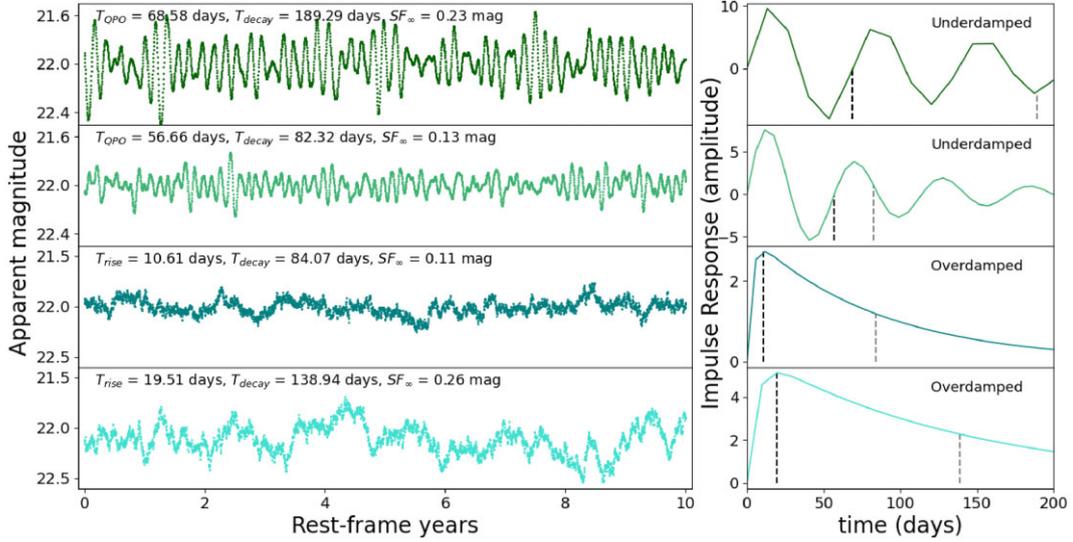

**Figure 4.** Four light curves simulated by DHO. Upper two are underdamped and lower two are overdamped cases, followed with their IR. For underdamped IR, the black dash line means $\tau_{\text{QPO}}$ and grey line means $\tau_{\text{decay}}$. In overdamped IR, the black line corresponds to $\tau_{\text{rise}}$, and grey $\tau_{\text{decay}}$.

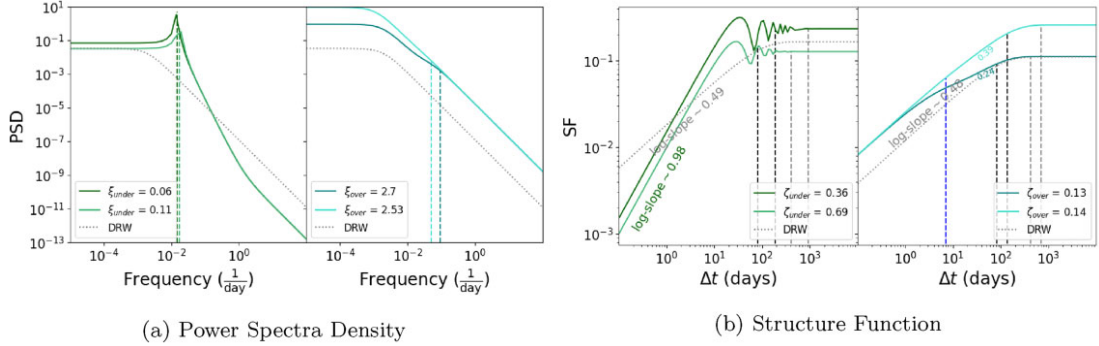

    (a) Power Spectra Density                                  (b) Structure Function

**Figure 5.** PSD and SF for four DHO light curves with underdamped and overdamped cases. The left-hand side (LHS) and the right-hand side (RHS) of each subplot represent the PSD and SF for underdamped and overdamped DHO, respectively. For Fig. 5a, $\xi$ is the damping ratio. The vertical dash lines represent $1/\tau_{\text{QPO}}$ (LHS) and $1/\tau_{\text{rise}}$ (RHS), respectively. For Fig. 5b, $\zeta$ means the ratio of $\tau_{\text{QPO}}/\tau_{\text{decay}}$ (LHS) and $\tau_{\text{rise}}/\tau_{\text{decay}}$ (RHS). The black and grey dash lines depict $\tau_{\text{decay}}$ and $5\tau_{\text{decay}}$, respectively. It is worth noting that when $\Delta t$ is equal to $5\tau_{\text{decay}}$, the SF has been steadily close to $SF_\infty$. The blue dash line on the right plot represents $\tau_{\text{blue}}$, where the overdamped SF slopes just decrease (Moreno et al. 2019).

**Table 3.** The ratios are the mean values of ($SF_\infty$ or $\tau$ in a given band) over mean ($SF_\infty$ or $\tau$ in the $u$ band) for the Stripe 82 quasar data.

| Band | $u$ | $g$ | $r$ | $i$ | $z$ | $y$ |
|---|---|---|---|---|---|---|
| $SF_\infty$ ratio | 1 | 0.88 | 0.75 | 0.66 | 0.63 | 0.61 |
| $\tau$ ratio | 1 | 1.29 | 1.51 | 1.28 | 1.26 | 1.26 |

Document). The baseline design elements for the WFD are: (i) cover at least 18 000 deg$^2$; (ii) average 825 visits per field, in all filters, over 10 yr, and (iii) obtain same-night, same-field revisit 'pairs'.

Within these bounds, several key characteristics that will define the main WFD survey remain to be determined, including: How the survey area is defined (`footprint`); how often each WFD field should be revisited (`cadence`) – both for intra- and inter-night visits; and what are the optimal filter distributions for the WFD fields (`filter`) and the optimal intra-night filter pairs for WFD revisits (`colours`).

## 3.1 LSST cadence strategies

The LSST is a complex survey with numerous science drivers. To this end, nearly 200 simulations of observing strategy have been generated that look into how different observations will drive the science goals and various metrics (see e.g. Lochner et al. 2021).

The survey strategies are summarized online[13] and we use the v1.7 version.[14] There are 'families' of observing strategies, with members of each family having related traits, e.g. the `visit_time` family are simulations examining the effect of the length of the individual visits, and the e.g. `u_long` family are simulations bearing on the length of the $u$-band exposure time.

The LSST OpSim is designed to simulate LSST observing strategies over the 10-yr survey, and different strategies have been tested to consider different scientific requirements (LSST Science Collaboration et al. 2017). Since this work is concerned with the global performance of ML neural networks in recovering quasar light-curve parameters, we decided to focus on five strategies (out of the over 190 available): a baseline and four different families. We do note that the LSST collaboration readily provides the cadence simulations, and we are providing our analysis codes, so studies with other observing strategies are easily produced.

**Table 4.** Brief overview of the five Legacy Survey of Space and Time (LSST) cadence strategies used in our study. Details are taken from the Jupyter Notebook [nbviewer.jupyter.org/github/lsst-pst/survey_strategy](nbviewer.jupyter.org/github/lsst-pst/survey_strategy).

| | Area with >825 visits | Unextincted area | Nvisits total | Median | Median Nvis | | | | | | Briefly desc |
|---|---|---|---|---|---|---|---|---|---|---|---|
| | | | | | *u* | *g* | *r* | *i* | *z* | *y* | |
| `baseline` | 17 982.70 | 15 174.43 | 2045 493.0 | 888.0 | 55.0 | 79.0 | 189.0 | 190.0 | 170.0 | 180.0 | Baseline |
| `u_long` | 18 112.80 | 15 011.61 | 1986 422.0 | 915.0 | 51.0 | 76.0 | 183.0 | 184.0 | 166.0 | 175.0 | *u* 1x60s |
| `filterdist` | 19 886.22 | 14 974.68 | 2221 366.0 | 1057.0 | 166.0 | 91.0 | 206.0 | 205.0 | 188.0 | 195.0 | *u* heavy |
| `cadence_drive` | 17 996.97 | 14 996.50 | 2046 411.0 | 893.0 | 46.0 | 115.0 | 194.0 | 179.0 | 160.0 | 170.0 | Add *g*, limit 200/night, contiguous |
| `rolling` | 17 960.88 | 15 051.89 | 2048 229.0 | 889.0 | 54.0 | 78.0 | 189.0 | 190.0 | 170.0 | 181.0 | 0.9 strength, 3 band |

The five survey strategies we focus on are:

(i) `baseline_nexp2_v1.7_10yr`, which we call baseline for short. This is the baseline WFD footprint, with the default observing behaviour having visits of $2 \times 15$-s exposures.

(ii) `u_long_ms_60_v1.7_10yr`, which we call *u*-long for short. Observations in the *u* filter are taken as single snaps, and we test increasing *u*-band exposure times. The cadence we choose has a $1 \times 60$ s, with the number of *u*-band visits left unchanged, resulting in a shift of visits from other filters to compensate for the increase in *u*-band observing time. With quasars being bright in the UV/optical, we are keen to see if additional *u*-band exposure improves the recovery of light-curve parameters.

(iii) `filterdist_indx4_v1.5_10yr`, which we designate filterdist for short. The aim is to evaluate the impact of changing the balance of visits between filters, where again we choose a '*u*-band heavy' cadence as our focus is on AGN.

(iv) `cadence_drive_gl200_gcbv1.7_10yr`, which we designate cadence_drive for short. This investigates the impact of reducing the gaps between *g*-band visits over the month, essentially down-weighting the lunar cycle. This aims to avoid long gaps in *g*-band coverage with the goal to improve transient discovery and variable characterization for longer time-scale objects which require bluer filter coverage (such as AGN and Supernovae). Our chosen cadence has 200 fill-in *g*-band visits each night in a contiguous area.

(v) `rolling_scale0.9_nslice3_v1.7_10yr`, which we designate rolling for short. A rolling cadence is where some parts of the sky receive a higher number of visits during an 'on' season, followed by a lower number of visits during an 'off' season. During the first 1.5 yr and the final 1.5 yr of the WFD survey, half the sky is covered uniformly, allowing for better proper motion coverage. This leaves 7.0 yr for 'rolling' observations, with the benefit that transient and variable phenomena are better observed, at the cost of each of the middle 7 yr will have no uniform survey coverage. Full details of the rolling cadences are given in Yoachim (2021).

We summarize the salient details of these strategies in Table 4.

### 3.2 Light curve with five cadence strategies

We selected and tested the LSST cadence simulations on the SciServer and used the Metrics Analysis Framework (Jones et al. 2014), to analyse the OpSim-simulated surveys. We chose several sky positions in WFD fields, and assume that each position corresponds to a quasar object. Given a set of DRW/DHO parameters, a mean magnitude, a chosen band, and an LSST cadence, we provide a `QuasarMetric` to return a realistic LSST-cadence light curve with MJDs, magnitude, and magnitude error. This metric can be used for any WFD survey strategy, so will be beneficial for future AGN light-curve analysis.

Fig. 6 presents a DRW-simulated light curve as observed under five cadence strategies. In the next section, we will discuss the influence of different strategies on quasar modelling with GPR and ML.

## 4 STOCHASTIC RECURRENT NEURAL NETWORKS

In this section we give a very high-level theoretical outline of SRNNs. We describe the SRNN architecture for our investigations, and how we implement the SRNN in practice. We note Fabius & van Amersfoort (2015), Chung et al. (2016), Fraccaro et al. (2016), Schmidt & Hofmann (2018), and in particular the notation from Yin & Barucca (2021) – as given in Table 5 – as important influences in what follows.

### 4.1 SRNN high-level overview

Derived and inspired from Bayer & Osendorfer (2015), Fraccaro et al. (2016) propose the idea of propagating stochasticity in a latent state representation with RNNs. They stack an SSM on deterministic RNNs to achieve a stochastic and sequential generative model and a structured variational inference network, which produce the output sequences and provide the model's posterior distributions, respectively.

This algorithm is particularly suitable for CARMA modelling as there are stochastic features, and CARMA can be represented as a format of SSMs. As such, here we applied SRNN to ingest AGN light curves with different LSST cadences and bands, and output modelled light curves with denser observations. The implementation of our SRNN, with the generative model and the inference model that we use in our study, is outlined in Fig. 7.

#### 4.1.1 Generative model

The role of the generative model is to establish probabilistic relationships between the target variable $y_t$, the intermediate variables of interest ($h_t, z_t$), and the input $x_t$. Within the generative model, a key part of how the RNN becomes an SRNN is the SSM. Inside a 'classical' RNN, the evolution of the hidden states $h$ is governed by $f$, a non-linear transition function: $h_{t+1} = f(h_t, x_{t+1})$, where $x$ is the input vector. For an SSM however, the hidden states are assumed to be random variables, $z_t$. In our model, the SSM layer latent states are Gaussian distributions. The input of the next layer is randomly sampled from these distributions, thus providing the stochastic features. The key design of our model is to stack an SSM layer on the last GRU $\boldsymbol{h}$ layer (Fig. 7, left-hand panel).

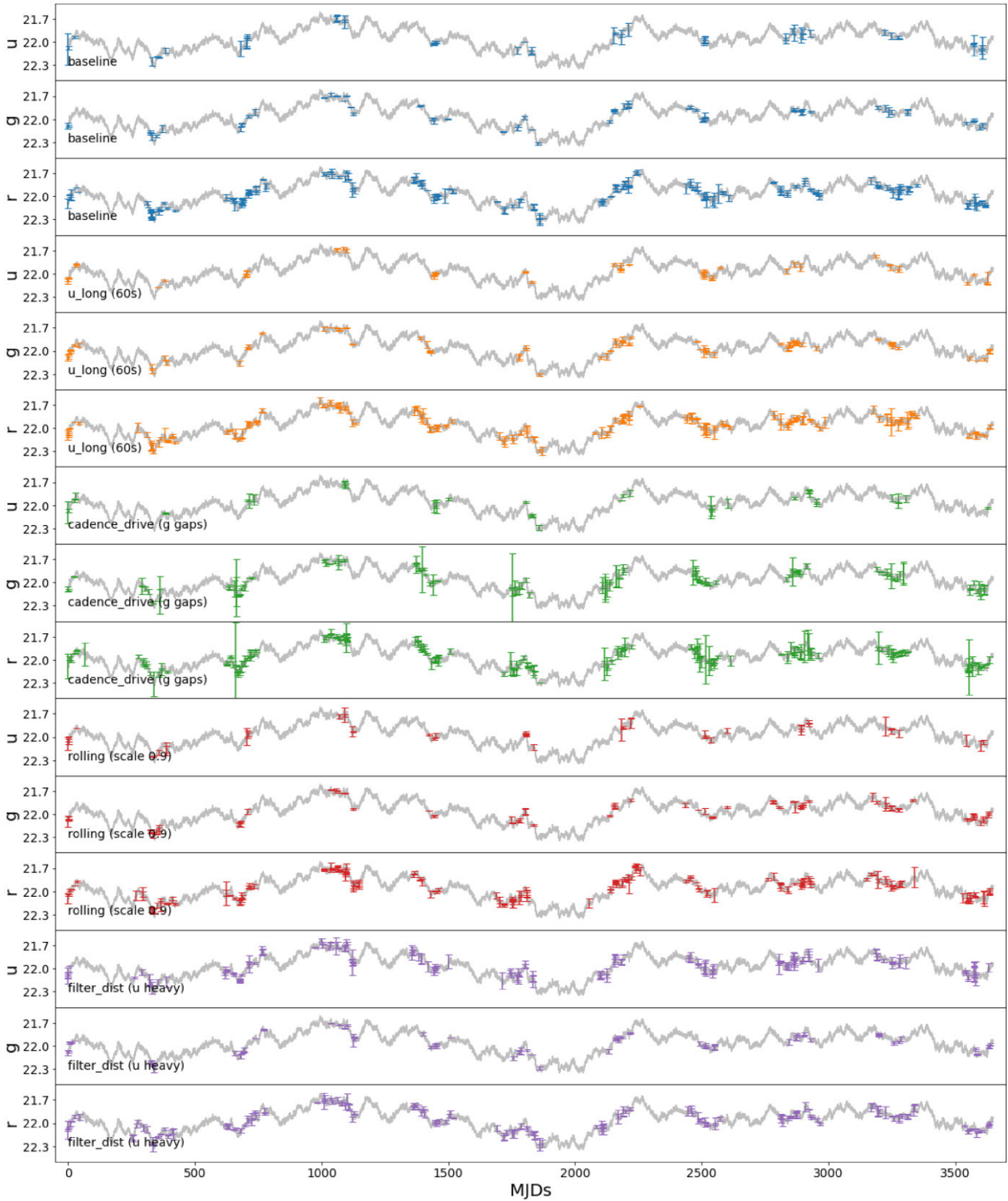

**Figure 6.** Identical DRW-simulated light curve ($SF_\infty = 0.17$, $\tau = 100$, and $z = 0$) at position RA = 0, Dec = -10, under five different cadence strategies. The grey light curve is the original one with dense observations. For each cadence strategy, $u$, $g$, and $r$ bands' cadences are selected. For u_long strategy with 60-s exposure time in $u$ band, the photometric error is much smaller than others. However, cadence_drive bearing reducing $g$-band gaps, some of its observations have large noise. The rolling cadence with scale 0.9 shows apparent trend in seasons after around 1000 d – the number of observations in the prior season grows and downs in its next season. Compared with u_long, filterdist for $u$-band heavy has more overall visits in $u$ band.

**Table 5.** Stochastic Recurrent Neural Network (SRNN) nomenclature and notation.

| Parameter | Description |
|---|---|
| $g_{\phi_a}$ | Represents one Gated Recurrent Unit (GRU) neuron function in general in the '$a$' layer |
| $\boldsymbol{h}_t$ | Hidden state |
| $\boldsymbol{x}_t$ | Input data |
| $\boldsymbol{x}_{1:T}$ | $= \{\boldsymbol{x}_1, \boldsymbol{x}_2, ... \boldsymbol{x}_T\}$ is a temporal sequence |
| $\boldsymbol{y}_t$ | Target variable |
| $z_t$ | Latent random variable |
| $\theta$ | Parameters set of $\{\theta_y, \theta_h, \theta_z\}$ |
| $p_\theta$ | Generative distribution |
| $q_\phi$ | Inference distribution |
| $\mathbb{E}$ | Expected value operator |
| $\mathcal{L}$ | Marginal log-likelihood function |
| $\mathcal{F}$ | Evidence Lower Bound (ELBO) |
| $D_{\mathrm{KL}}$ | Kullback–Leibler divergence (KL divergence) |

Given the fact that CARMA can be written as developing Gaussian distributions over time-steps, the SSM layer is expected to present similar functions for generating the output sequences. In this way, the model can learn both the long-term dependency within a sequence, as well as the stochastic features of the input sequence. Combining the non-linear gated mechanisms of the RNN with the stochastic transitions of the SSM creates a sequential generative model that is more expressive than the RNN and better capable of modelling long-term dynamics than the SSM. Fig. 7a shows the architecture of the generative model.

### 4.1.2 Inference network

The second part of the SRNN is the inference network. Here, the prior distributions in the generative model learn from the posterior distributions by KL divergence (Kullback & Leibler 1951). While training, the generative model and inference network are both implemented, and learn from the backpropagation.

Fig. 7b outlines the inference model architecture. For each step, latent states $a_t$ are fed with the combinations of $\boldsymbol{y}_{1:T}$ and $\boldsymbol{h}_{1:T}$, so it has the information of the future time-steps from target sequences as well as previous steps' information from the last $h$ layer in the generative model. In order to let posterior distributions contain the information from the future steps, the RNN layer is reversed in the time dimension.

### 4.1.3 Loss function

Initially, we wish to maximize the marginal log-likelihood function $\mathcal{L}$, where $\mathcal{L}(\theta) = \sum_i \mathcal{L}_i(\theta) = \log p_\theta(y_{1:T}|\boldsymbol{x}_{1:T}, z_0^i, d_0^i)$. However, the random variable $z_t$ in the non-linear SSM cannot be integrated out analytically (see e.g. Kingma & Welling 2014, section 2.1). Therefore, we instead aim to maximize the ELBO (also known as the variational lower bound) given as $\mathcal{F}$, with respect to the generative model parameters $\theta$ and an inference model parameter, $\phi$. Thus, the objective function of our SRNN is $\mathcal{F}(\theta, \phi)$, given in Appendix B1. In practice, minimizing the loss function is more intuitive and convenient for implementation. Therefore, we present our loss function equation (7), which can be derived from ELBO.

$$\mathcal{L} = NLL + D_{\mathrm{KL}_i}\left(Q(z) \parallel P(z)\right)$$

$$= \frac{1}{N_{\mathrm{lc}}} \sum_i^{N_{\mathrm{lc}}} \sum_t^{N_{\mathrm{T}}} \Bigg( \left[ (m_{\mathrm{target}_t} - m_{\mathrm{rec}_t})^2 - m_{\mathrm{error}_t}^2 \right]$$

$$+ \log \frac{\sigma_{p_i}}{\sigma_{q_i}} + \frac{\sigma_{q_i}^2 + (\mu_{q_i} - \mu_{p_i})^2}{2\sigma_{p_i}^2} - \frac{1}{2} \Bigg). \tag{7}$$

This is formed from the negative log likelihood and KL divergence ($D_{\mathrm{KL}}$; Kullback & Leibler 1951) of the prior and posterior distributions of $z$. $Q$ and $P$ correspond to the approximate posteriors in the inference network and priors in the generative model, respectively. Furthermore, the magnitude errors are considered in the loss calculation. $N_{\mathrm{lc}}$ is the number of light curves per loss calculation. $N_{\mathrm{T}}$ means the length of each light curve. $m_{\mathrm{target}, t}$ means the $t$-th observation of the target light curve, and $m_{\mathrm{rec}_t}$ means the $t$-th observation of the reconstructed light curve by SRNN. $m_{\mathrm{error}_t}$ represents the magnitude error in the $t$-th observation of the LSST-cadence light curve (also the input light curve), with 0 for vacant observations.

### 4.2 Parameter configuration

Our SRNN is built using the open-source software ML library TensorFlow (tf), with Keras as the backend. We used Google Colab as the computing platform, and chose the GPU P100 nodes option.

For the generative model, we set one bidirectional GRU $\boldsymbol{h}_t$ layer with 32 neurons, followed by two bidirectional GRU layers – for the $\mu_z$ and $\log \sigma$ priors – with 32 neurons for each layer. All layers apply the default hyperbolic function tan$h$ as the activation function. We add a 'bidirectional' wrapper to each GRU layer as we found during testing it increased the accuracy of the light-curve reconstruction. For the inference network, similarly, there are three bidirectional GRU layers with 32 neurons, corresponding to $\boldsymbol{a}_t$, the posterior $\mu_z$, and $\log \sigma_z$ respectively. The dimension of the inputs includes magnitudes $y$ (and magnitude errors $y_{\mathrm{err}}$ for DRW and DHO overdamped cases). The outputs are the dense light curves, with the observation length identical to the input sequences.

As the goal of applying SRNN is to model the whole light curves, including those dates where there are no observations, it is necessary to discuss how to deal with these vacant time-steps. We cannot impute these values using Autoregressive integrated moving average (ARIMA), (e.g. Saputra et al. 2021) and Neural-Networks-techniques (e.g. Li et al. 2020; Shu et al. 2021), as the SRNN would not then need to learn how to fill the gaps. Our solution is: For each dimension, the value is first reduced by its mean values, and then zeros are added for those time-steps where the observations are vacant. In other words, we pre-processed those vacancies with the mean values. As RNNs do not allow 'NaN' values included in inputs, adding zeros can provide these vacancies with initial values. During training, SRNN is expected to learn the correlations between steps and predict the corresponding values for these vacancies. As such, adding zeros has less physical meaning, but simply satisfies the RNN input rules. However, when the vacant period is too long, the predicted values will move close to the mean values.[15]

---

[15] We also test to fill in these vacancies with other values. For example, making a straight line between neighbouring observations, starting with the previous observation and ending with the latter one, and then filling in the vacancies with the corresponding values on the line. However, this harms SRNN's predictions for their true values; adding a masking layer after the

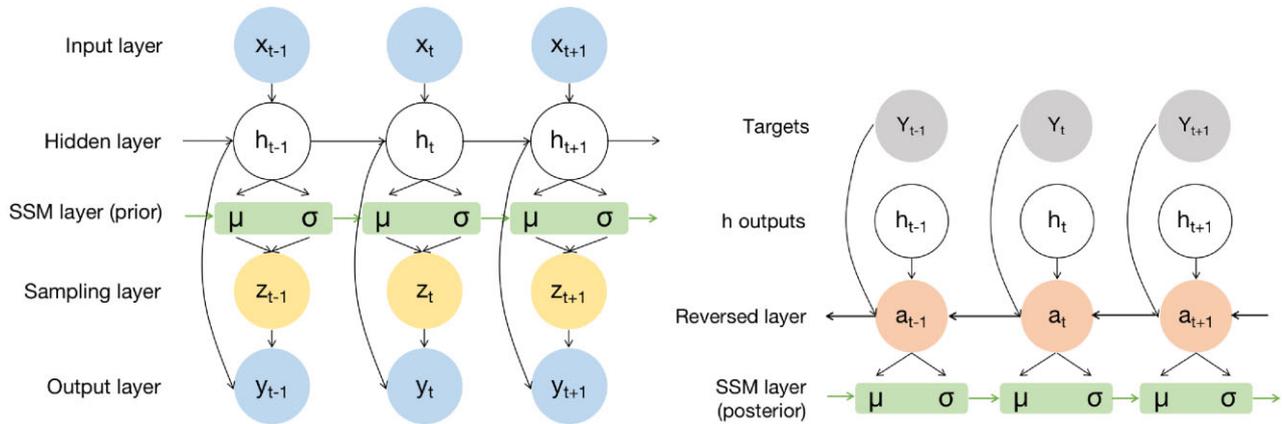

**Figure 7.** (a) The architecture of the generative model (left-hand panel). The inputs layer is connected with a hidden layer (called *h*), which can be a GRU or LSTM layer. There could be multiple *h* layers, and in this diagram we only show one. An SSM layer is interlocked with the last *h* layer, represented by prior Gaussian distributions $N_{prior}$ in the diagram. Note that the SSM layer is implemented by two GRU layers in the programs, producing $\mu_{(p)}$ and $\log \sigma_{(p)}^2$, respectively. A sampling layer is used for randomly sampling a variable from each Gaussian distribution at each step. The output of the sampling layer is further combined with the output of the last *h* layer, and passed to the output *dense* layer, to generate output sequences. (b) Inference network. This is for providing the posterior Gaussian distributions. The target output sequences are combined with the ***output*** of the last *h* layer (also in the generative model), and they are delivered to a reversed GRU layer, called *a*. In this way, the output of the *a* layer at each step contains the information both from its future steps (from the target) and its previous step (from *h*). Furthermore, an SSM layer is stacked on the *a* layer, producing the posterior $\mu_{(q)}$ and $\log \sigma_{(q)}^2$ for each step, which is compared with the priors by calculating the KL divergence. During training, this divergence is considered and works with the negative log likelihood to minimize the whole loss function.

We chose the Adam Optimizer (a stochastic gradient descent method derived from adaptive moment estimation) with a learning rate $r = 5 \times 10^{-3}$, $\beta_1 = 0.900$, $\beta_2 = 0.999$, and $\epsilon = 1 \times 10^{-7}$. The batch size is 128 per training epoch, and the maximal training epoch number is 300. `tf.keras.callbacks.EarlyStopping` is also applied with the patience set to 3 in order to stop training if the loss function is no longer decreasing.

### 4.3 Data set and training plan

The outputs from the SRNN are light curves. To see how the SRNN performs in 'ideal' circumstances, we input a dense and uniformly sampled light curve with 3650 points (a daily observation for 10 yr) to the SRNN, and target a similarly dense light curve as the output. The training data involve 7200 light curves, and test data involve 1800 light curves.

To predict the light curve during seasons in the 10-yr LSST-simulated light curves, and see how SRNN could help recover the CARMA DRW and DHO parameters, light curves based on the five considered LSST cadences are also fed as the input of the SRNN, and dense light curves will be the targets. The SRNN is expected to learn the trends between seasons. In this instance, the training data involve 14 400 light curves, and the test data are 3600 light curves.

We also want to predict future observations based on observed light curves, and again see how the SRNN could help recover DRW and DHO parameters for light curves with shorter duration. As such, we shortened the length of dense and LSST-cadenced light curves to 6–9 yr, and let SRNN to predict the next 1–4 yr's light curves.

For both the LSST light curves as well as the SRNN predicted daily light curves, we apply the GPR method (Section 2.2.3) to measure input layer is also tested, in order to mask all the missing values with zeros and ignore them by deactivating their passing neurons while training. This design dramatically slows down the training process, which is about 5 times longer on GPU cores, and the predictions are not as good as the architecture without masking.

CARMA parameters. We note, GPR is only one method to estimate the CARMA parameters of the light curves; it is not as accurate as Markov chain Monte Carlo (MCMC), but more efficient.

## 5 RESULTS

After generating the model light curves in Section 2 and simulating observations through the LSST cadences as discussed in Section 3, we use the SRNN described in Section 4 to reconstruct the full light curves and calculate CARMA model parameters. In this section, we report the results.

### 5.1 Light-curve modelling

#### 5.1.1 Full and uniform seasonal light-curve modelling

Fig 8 shows results of SRNN modelling for one DRW light curve (SF$_\infty$ = 0.1 mag and $\tau$ = 307 d) under different samplings. We test four example samplings: The light curve can be fully sampled, or have observation gaps of 30, 60, or 120 d in-between 30 d of observations. Mean Absolute Error (MAE) is presented for qualifying the influence of the increasing gap length on the whole light-curve modelling.

These examples illustrate that the longer the gaps in observations, the lower the accuracy of the model and the larger the MAE. From the plot it can be seen that the SRNN is able to predict the overall trend of each gap, which is similar to its real values. It also can simulate the stochastic characteristics of the sequence, by randomly sampling from Gaussian priors (the mean and standard shown in lower two panels) for each step.

#### 5.1.2 LSST light-curve modelling

For LSST-cadence light curves, it is harder for the SRNN to learn the correlations between each time-step, as the cadences are often not uniform and have longer intervals. The SRNN is trained for

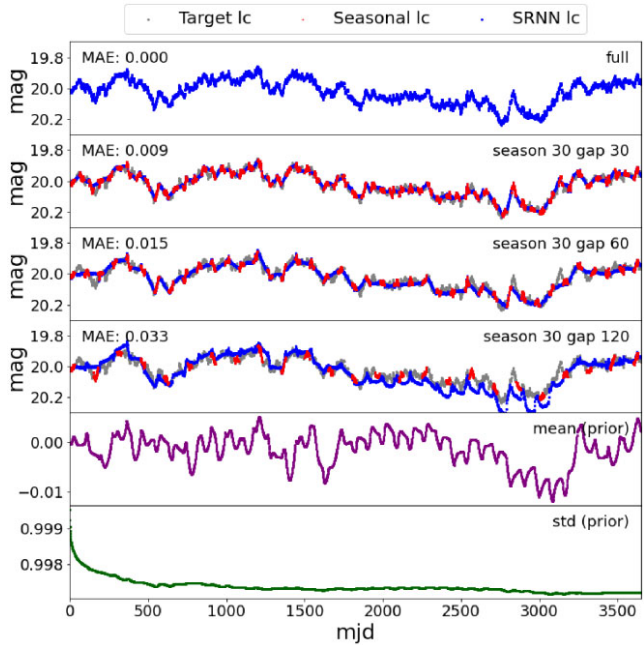

**Figure 8.** Multiple SRNN modelling for one DRW process ($SF_\infty = 0.1$ mag and $\tau = 307$ d) with different uniformly designed cadences. In this experiment, the magnitude errors are set to zero for simplicity. the upper five rows correspond to input light curves with different cadences, which is labelled on the top right of each panel, respectively. The input light curves are shown in red, modelled light curves by SRNN are shown in blue, and the grey means the full light curves (10 yr's observation length; one observation per night). The Mean Absolute Error (MAE) between the model light curve and the target light curve is calculated for each case and displayed in the upper left corner. The last two rows show the mean and standard deviations of the Gaussian prior distributions used by one example neuron (of the many that contribute in the SSM layer, from which the random variable $z_t$ is sampled (Fig. 7). The case shown is for the input light curve with 'season 30, gap 120' cadence.

the different models (DRW and DHO underdamped/overdamped) separately.[16]

Fig. 9 presents the reconstructions of a quasar with a DRW CARMA process, given in the $r$ band with the five LSST cadences. The input DRW parameter values are $SF_\infty = 0.218$ mag and $\tau = 691.656$ d, and redshift $z = 1.677$. The input light curve is the same for all five cadences.

Fig. 10 presents the reconstructions of a quasar with a DHO CARMA process, in the overdamped case. The input DHO parameter values are $SF_\infty = 0.215$, $\tau_{rise} = 11.208$ d, $\tau_{decay} = 1118.304$ d, and redshift $z = 1.233$.

The modelling of these two cases shows that SRNN is able to predict the gaps between seasons with stochastic characteristics in general, though it fails to predict accurately the specifics of the variability during the gaps. This is unsurprising, as the CARMA process at each step is indeed a (Gaussian) random variable.

Fig. 11 shows the reconstruction of a quasar with a DHO CARMA process in the underdamped case. The input DHO parameter values are $SF_\infty = 0.236$, $\tau_{QPO} = 44.193$ d, $\tau_{decay} = 239.414$ d, and redshift $z = 1.372$. This plot shows that SRNN is better at learning the periodicity characteristic of the DHO-underdamped process. With few observations, SRNN can predict the vacant observations well.

---

[16]For input light curves with few observations, it is hard for the SRNN to figure out which case it belongs to.

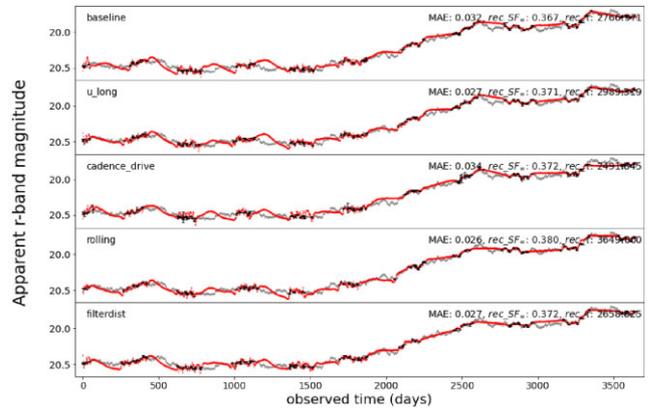

**Figure 9.** The reconstructions of a quasar with a DRW CARMA process, given in the $r$ band with the five LSST cadences. The original DRW parameter values are $SF_\infty = 0.218$ mag, $\tau = 691.656$ d, and redshift $z = 1.677$. The input light curves are the observed light curves with time dilation considered, which are identical for all five cadences, presented with grey points. Then they are sampled with different cadences respectively, shown in black points with error bars. The reconstructed light curves (by SRNN) are shown with red points. The recovered reconstructed parameters (observed) are given in the top right of each panel. MAE represents the Mean Absolute Error, rec_$SF_\infty$ and rec_$\tau$ correspond to the parameter estimation (by GPR) after reconstruction, which are shown on the top right of each panel.

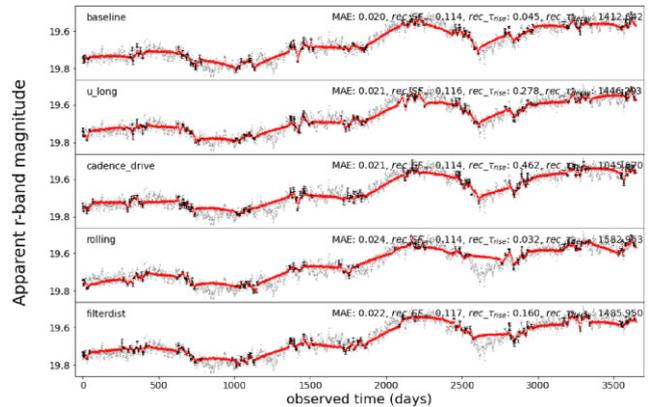

**Figure 10.** The reconstructions of a quasar with a DHO CARMA process, in the overdamped case. Magnitudes are reported in the LSST $r$ band with the LSST cadences from top to bottom panel. The input DHO parameter values are $SF_\infty = 0.215$, $\tau_{rise} = 11.208$ d, $\tau_{decay} = 1118.304$ d, and redshift $z = 1.233$. The reconstructed (observed) DHO parameters are given in the top right of each panel.

### 5.1.3 Problem of 'filling the gap'

Here, we particularly discuss how SRNN modelling fills in the gaps between distant observations. As can be seen from Figs 8–11, SRNN can reconstruct the input observations when gaps are reasonably short compared to the time-scale of variability, but for large gaps, SRNN's general performance is weak. The following factors all affect the SRNN light-curve reconstruction:

(i) Number of observations: Unsurprisingly, the number of provided observations is a major determinant how much information that SRNN can digest.

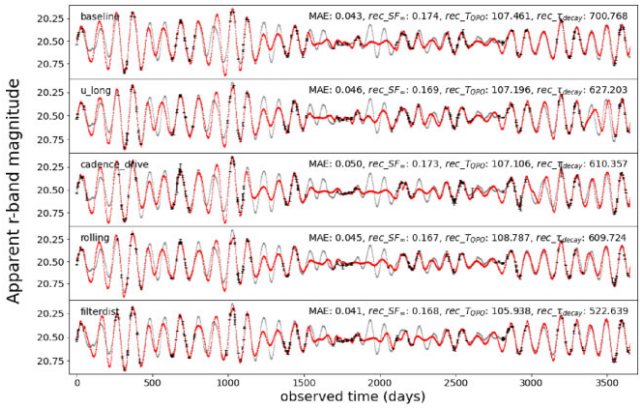

**Figure 11.** The reconstructions of a quasar with a DHO CARMA process, in the underdamped case. The input DHO parameter values are $SF_\infty = 0.236$, $\tau_{QPO} = 44.193$ d, $\tau_{decay} = 239.414$ d, and redshift $z = 1.372$ The reconstructed (observed) DHO parameters are given in the top right of each panel.

(ii) Cadence strategies and different bands: Although different cadences with the same total number of observations in a given band give similar results, in reality different cadences have different allocations of observations to individual bands. Those in which a higher proportion of observations are allocated to a particular band lead to better SRNN modelling results in that band (at the expense of other bands). On the other hand, regular samplings (e.g. Fig 8) also can improve the predictions of gaps. More investigations are shown in Section 5.2.

(iii) Level of perturbations: high $SF_\infty$/short $\tau$: Light curves with extremely high variability or very short time-scales are difficult for SRNN to model, since the limited number of observations and long gaps are not sufficient for SRNN to learn these features.

(iv) Quasi-periodicity: Fig 11 shows that SRNN is better at modelling light curves with quasi-periodic features. It can be seen that for these light curves with durations 3500 d and quite different cadences, SRNN is able to predict the observations well, and even only several observations can help SRNN to greatly recover the trends with high accuracy.

(v) Assumption of stationarity: All simulations and fittings in this paper are based on the stationary model CARMA, though the real AGN variability could be non-stationary (Tachibana et al. 2020). The main reason for our assumption of stationarity is that at this stage, not enough real and good-quality AGN light curves are provided for the training set (for both inputs and targets), while CARMA (especially DRW) has been a popular model for AGN variability study for years. It is the closest and simplest model that could help to achieve AGN light-curve simulations, though it does have drawbacks and discrepancies compared with the real ones. From this perspective, SRNN should not be expected to recover the real short-term events happening in gaps, as CARMA light curves are not generated by deterministic physical processes.

To summarize: For the LSST cadences shown in Figs 9–11, long gaps exist between observations, and for the reasons above, the SRNN model struggles to impute the behaviour during these gaps, especially for the non-periodic DRW and DHO-overdamped cases. This will turn out to be an important limitation when attempting to infer CARMA parameters from these light curves.

## 5.2 Parameter estimation analysis

The main motivation of our paper is to investigate whether using an ML algorithm on (synthetic) LSST quasar light-curve data would be able to detect and/or mitigate any biases in derived CARMA model parameters. Specifically, we have modelled the CARMA(1,0) DRW and CARMA(2,1) DHO processes, the latter in both the overdamped and underdamped cases. Here, we report the results of these investigations.

### 5.2.1 Metric for CARMA parameters

We design a metric, $M_{err}$, for evaluating how the CARMA parameters are recovered by SRNN-modelled daily light curves, compared with LSST-cadence light curves. This metric is used for the comparisons between combinations of different cadences and bands, and it can also be used as an ensemble metric for each cadence with all bands considered.

$$M_{err} = \frac{1}{N} \sum_i^N \sum_j^M \frac{\left| \theta_{SRNN}^{(i,j)} - \theta_{LSST}^{(i,j)} \right|}{\theta_{in}^{(i,j)}}, \qquad (8)$$

where $N$ is the number of light curves used in evaluating the metric. $M$ is the number of parameters for the relevant CARMA model. $\theta_{SRNN}$, $\theta_{LSST}$, and $\theta_{in}$ are the parameters recovered (by GPR) on the SRNN reconstructed light curves, the parameters recovered from the LSST simulated light curves, and the input parameters, respectively. We calculated the absolute values of differences between $\theta_{SRNN}$ and $\theta_{LSST}$, and then divide them with $\theta_{in}$ in order to measure each parameter with the same scale. This metric can be extended to any LSST cadences and any bands.

Fig 12 shows our calculated metrics, for each of the five LSST cadences and six LSST bands, in the DRW, DHO-overdamped, and DHO-underdamped cases. We also show ensemble metrics for each cadence in each CARMA model. Before metric calculations, those objects with estimated $\tau_{decay}$ longer than $10^4$ d are regarded as outliers and removed.[17]

In this plot, it can be seen that the DHO-underdamped case always gains the lowest $M_{err}$, as the SRNN algorithm can better simulate this kind of light curve, followed by the DRW case. The DHO-overdamped case always has the largest $M_{err}$. Regardless of which CARMA model is used, the $u$-band light curves usually have the largest $M_{err}$. For most bands, filterdist generally has the lowest (best) $M_{err}$, which shows that the number of observations plays a key role in parameter estimation.

To make this more explicit, the lower right panel of Fig 12 shows ensemble metrics for each cadence. Here, three conditions are considered. The mean metric value from all LSST bands in a given cadence shows that the differences between baseline, u_long and cadence_drive are tiny for the DHO-underdamped and DRW models. The filterdist cadence always gains the lowest mean $M_{err}$ among all cadences, with rolling having the largest value. Such results indicate that filterdist might be the optimal cadence for SRNN modelling for this case, and rolling is the worst option due to the long gaps in coverage.

However, as the large $M_{err}$ in the $u$ band, which is poorly sampled for most cadences, drives the mean metric to higher values, we also present a mean metric that considers only the five redder bands. In this case, cadence_drive gains the lowest metric. If we consider single-band modelling, taking only the minimal value (for any band, but typically r), u_long, cadence_drive, and filterdist all perform well, with the best cadence depending on the CARMA model used.

---

[17]LSST only has 10-yr observation length, $\tau$ much longer than 10 yr cannot be estimated correctly.

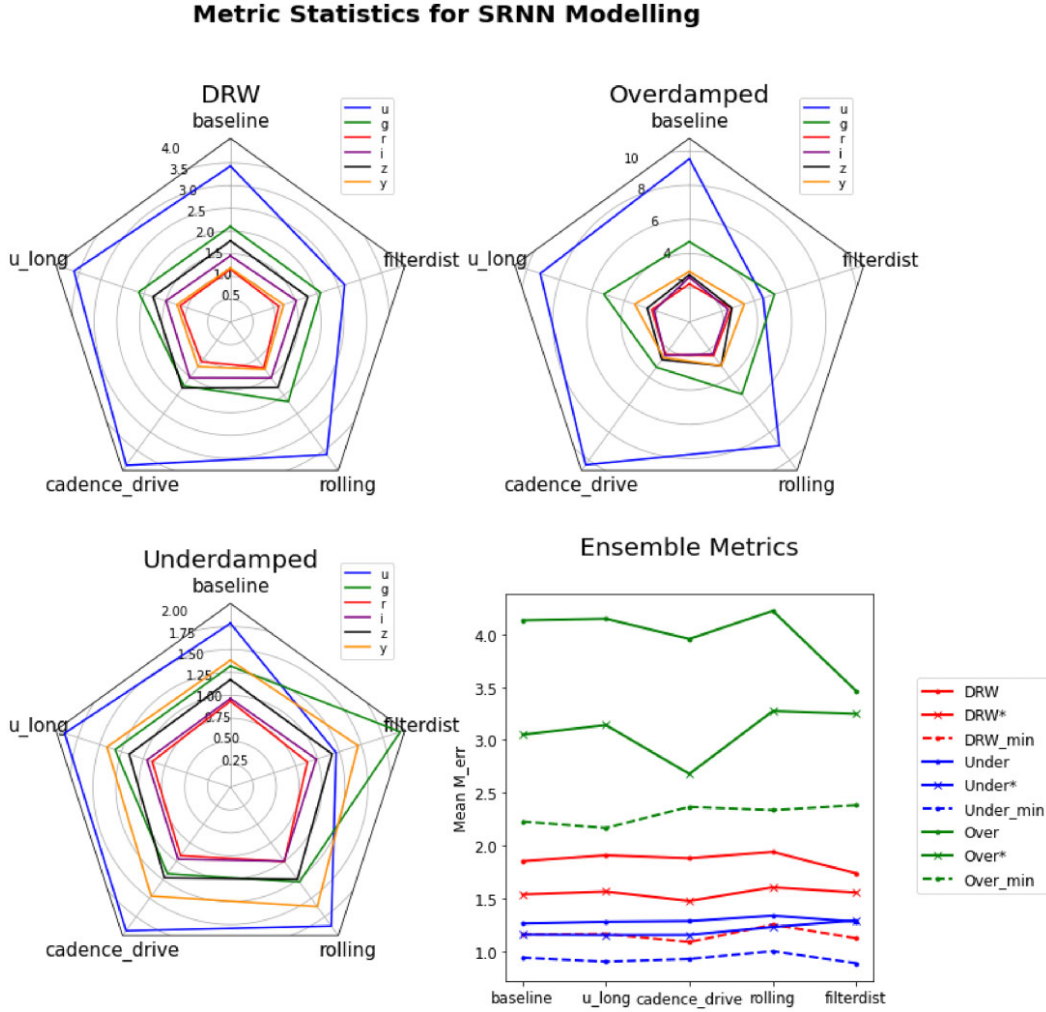

**Figure 12.** CARMA metric under different LSST cadences. Upper left, upper right, and lower left subplots show the metrics with DRW, DHO-overdamped, and DHO-underdamped cases, respectively. The lower right subplot shows the ensemble metric for different CARMA cases and cadences. The solid lines with round markers correspond to the mean value with each cadence; the solid lines with 'x' marker (the labels with *) correspond to the mean value without u band; the dashed lines correspond to the minimal value for any band (usually r) within each cadence.

*5.2.2 DRW parameters analysis*

Finally, we investigate the derived parameters and possible biases in SRNN modelling of AGN, concentrating on the DRW case as it is the most popular model used to analyse AGN time series. Analysis for DHO cases will be conducted in future work.

Fig. 13 shows the trend in recovered time-scales compared to the model inputs, expressed as log ($\rho_{out}$) and log ($\rho_{in}$), where $\rho_x \equiv \tau_x/(10$ yr) (the LSST survey length). For small time-scales, the recovered $\tau_{out}$ from the SRNN-modelled light curves are highly overestimated. This is because a shorter time-scale leads to more perturbations within a given gap. SRNN shows worse performance in predicting the highly variable magnitudes in longer gaps and fewer observations, resulting in relatively flat predicted light curves and overestimated time-scales. For longer time-scales, when log ($\rho$) ≥ −1, parameter recovery with the SRNN is somewhat better than using the LSST data alone, but the time-scales are underestimated in both cases. This is because an observing length longer than ∼8–10 times τ is required to provide sufficient information for unbiased τ estimation (Kozłowski 2017; Kozłowski 2021; Suberlak et al. 2021). Put simply, when τ is longer, e.g. 5 yr, and the observing length is 10 yr, there are not enough samples for calculating the magnitude differences at high Δt, resulting in underestimation of τ.

SRNN-modelled and LSST-cadence metrics almost overlap in r, i, z, and y bands. For u band, in all cadences, $\rho_{out}$ for the SRNN reconstructed light curves is always longer than that for LSST cadence light curves. Examining the ranges of log ($\rho_{out}$) for the six bands, the r, z, and y bands in all cadences allow the best performance: They are closer to the diagonal than other combinations. Especially for z band, some SRNN metrics at small time-scales are very similar to LSST ones. For a given band, the SRNN performance is similar across most cadences, but larger differences are seen in u band – in this case filterdist performs relatively better than the others.

Fig 14 shows how well $SF_\infty$ can be recovered from the SRNN-modelled and LSST light curves. It shows that $SF_\infty$ in all bands and cadences are underestimated. When log ($SF_{in}$) increases, log ($SF_{out}$) moves further away from the diagonal. The main reason is a positive correlation between $SF_\infty$ and τ (see Fig 1): High $SF_\infty$ often corresponds to longer time-scales, and such light curves require more observation time (time dilation is also considered) to reach the plateau of the SF plot (Fig 3). Given the 10-yr LSST survey length, SF$_\infty$ of many objects with long time-scales will be underestimated.

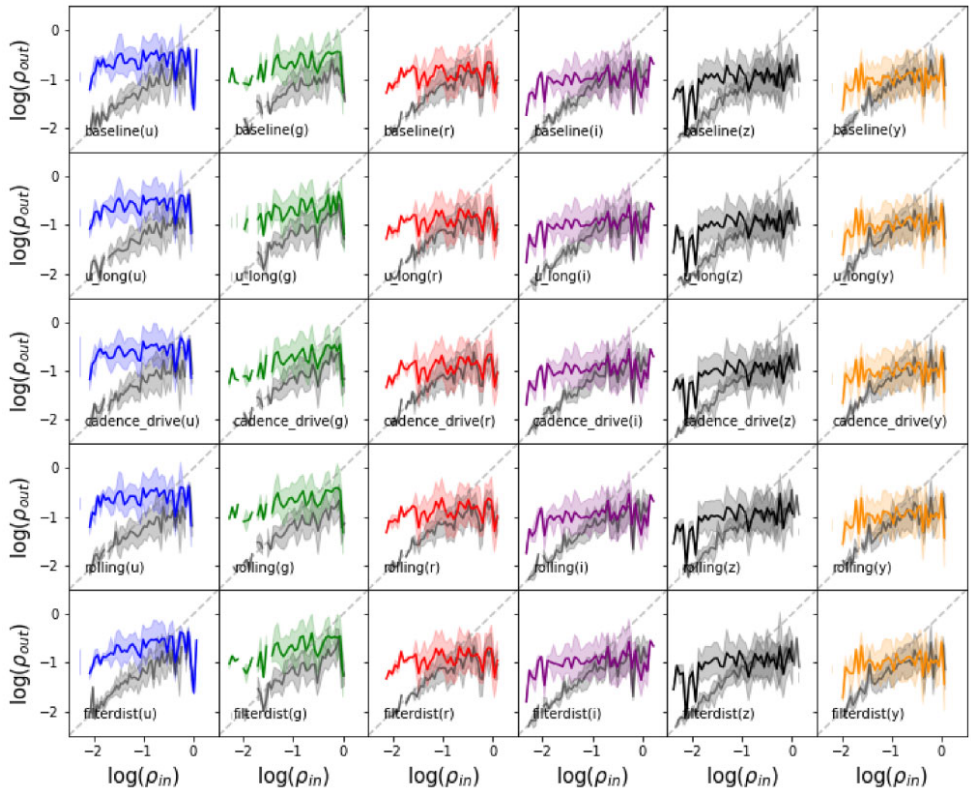

**Figure 13.** $\tau$ estimation for DRW processes. $\rho_{in}$ and $\rho_{out}$ are defined as $\tau_{in}$ (the input time-scale) and $\tau_{out}$ (the recovered time-scale from the simulated SRNN or LSST data) divided by 10 yr (the LSST survey observation length). Each row corresponds to a different LSST cadence, and each column corresponds to a band. Results from the SRNN-modelled light curves are shown with colours, and results from LSST-cadence light curves are shown in grey. The shaded areas indicate the standard deviation around the median, in 50 bins in $\rho_{in}$, following outlier rejection.

In general, the SF$_{out}$ distributions from the SRNN and LSST light curves overlap in most cases except for $u$ band. SF$_{out}$ estimations from $u$-band SRNN light curves, with the baseline, u_long, cadence_drive, and rolling cadences, are lower than those from LSST light curves. This is because light curves with sparse observations make it harder for SRNN to learn their features, resulting in smooth predictions during observing gaps. When more observations are allocated (such as in the filterdist cadence), SF$_{out}$ gets closer to the diagonal.

# 6 DISCUSSION

## 6.1 SRNN modelling performance

Our SRNN model shows the ability to reconstruct realistic AGN light curves for the three CARMA models of interest. However, we have also identified biases and limitations in this method. In particular, the relative inability to predict observations during long gaps means that modelling results are very sensitive to season length and variability time-scales, as illustrated in Figs 12–14.

The SRNN model clearly struggles to recover the damping time-scale, predicting $\tau \sim 1$–2 yr regardless of $\tau_{in}$ (Fig 13). The problem at short time-scales is due to the problem of gaps in cadence much longer than $\tau$, with the result that the SRNN representation does not capture the true variability in these gaps (Section 5.1.3). At longer $\tau_{in}$, both the SRNN and LSST-cadenced data return underestimates. This is because of the 10-yr survey duration: For $\tau \gtrsim 1$ yr, the data do not have a long enough baseline to give sufficient samples of $\tau$ (Kozłowski 2017).

The model performs better in recovering the SF. As shown in Fig 14, this can be recovered as accurately from the SRNN light curves as from the LSST light curves, in all bands except $u$, where the lower number of detections likely inhibits SRNN reconstruction. Interestingly, the recovered values of SF$_\infty$ are systematically slightly below the input values for both SRNN and LSST light curves.

## 6.2 Comparisons with previous work

The range of ML techniques is vast and here we concentrate on five main artificial neural network types: the RNN, the RAE, the VRAE, the VRNN, and the SRNN. We also include multiband GPR here as the motivation of Hu & Tak (2020) is consistent with this paper.

Very high-level descriptions of these five architectures and astronomical applications are given in Table 6. We also compare the motivations, input formats, architecture, and results of Tachibana et al. (2020), Sánchez-Sáez et al. (2021), Hu & Tak (2020), and this paper, shown in Table 7.

Our study represents the first application of an SRNN to astronomical research. The unique design of the SRNN is to stack a SSM layer on the traditional RNN layer in order to learn and produce both the underlying features and corresponding stochastic fluctuations. SRNN is suitable for modelling light curves with extremely high variability, especially AGN. Given an AGN, when dense observations and uniform cadences are provided, SRNN can learn its features better. However, SRNN also has some weaknesses. Its input and output sequence lengths should always be identical, and its forecasting ability is relatively weak since the behaviour is inherently random. Nonetheless, given its ability to reconstruct realistic AGN light curves, we encourage further investigations regarding its architecture.

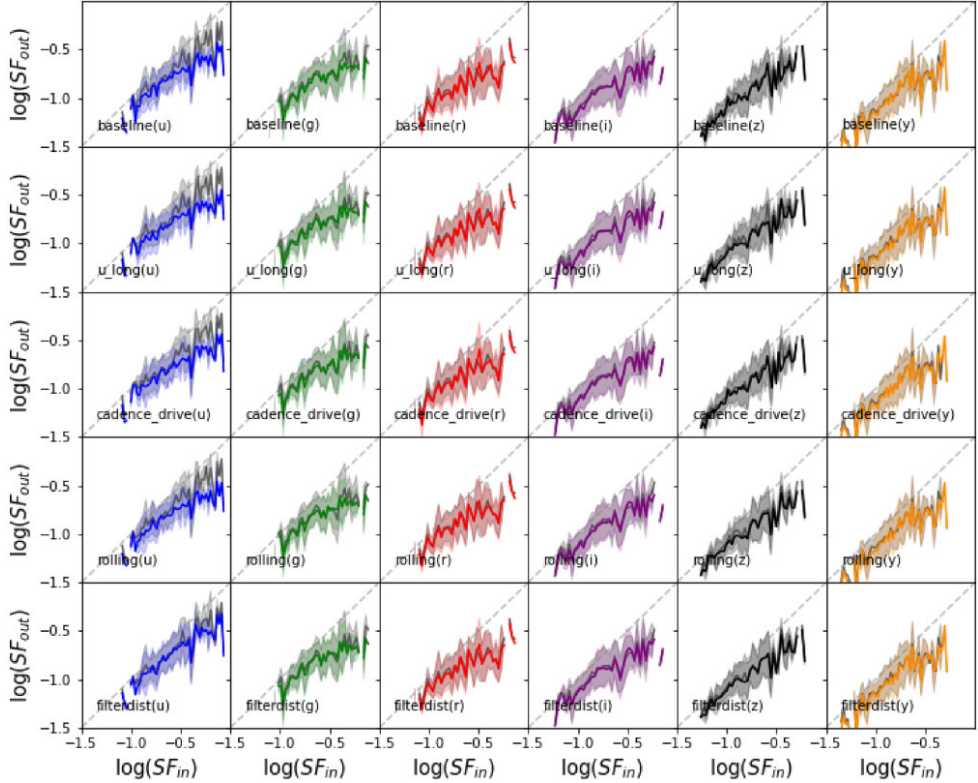

**Figure 14.** Same as Fig 13 but for $SF_\infty$. $SF_{in}$ and $SF_{out}$ are the real $SF_\infty$ and estimated $SF_\infty$ from SRNN or LSST light curves, respectively. The fitting lines with colours are for metrics of SRNN-modelled light curves, and the grey ones are for LSST light curves.

**Table 6.** A very high-level description of the four architectures most discussed in this paper.

| Machine Learning (ML) network | Description | Recent astrophysics studies |
| --- | --- | --- |
| Recurrent Neural Network (RNN) | RNN uses multiple layers of recurrent cells [RNN, GRU, Long Short-term Memory Unit (LSTM), etc.] for processing sequential data. Connections between nodes form a directed graph along a temporal sequence. | Dékány & Grebel (2020), Fremling et al. (2021), and Burhanudin et al. (2021) |
| Recurrent Auto-Encoder (RAE) | Its encoder learns a representation (encoding) for a set of data, typically for dimensional reduction (2D time sequences to 1D latent variables), by training the network to extract the inherent features from input sequences, and ignore insignificant or noisy data. Then, the representation, or so-called 1D latent variables are fed into a decoder, which decodes the features and generate output sequences. This architecture is design for sequence modelling and forecasting. | Naul et al. (2018), Tsang & Schultz (2019), Jamal & Bloom (2020), Tachibana et al. (2020), and Villar et al. (2021) |
| Variational Recurrent Auto-Encoder (VRAE) | Similar to RAE that it consists of an encoder that learns a mapping from input sequences to latent representation, and a decoder mapping from the latent representation to outputs. However, the variational approach maps the data to Gaussian distributions of latent variables instead of determined variables. Such design provides flexibility for VRAE to generate new and varying outputs with similar features. | Sánchez-Sáez et al. (2021) |
| Variational Recurrent Neural Network (VRNN) | VRNN contains a Variational Auto-Encoder (VAE) at each step, and each VAE (including the prior of the latent random variables) is dependent on the hidden state at the previous step $h_{t-1}$. | None |
| SRNN | Extended from VRNN, SRNN model combines both VRNN and State Space Model (SSM) advantages. Compared with VRNN generating deterministic values, SRNN provides stochastic random variables sampled from latent Gaussian distributions, thus more suitable for time-series study with high variability. | This paper |

**Table 7.** Comparison between ML applications to AGN in Tachibana et al. (2020), Sánchez-Sáez et al. (2021), Hu & Tak (2020), and this paper.

| Attribute | Tachibana et al. (2020) | Sánchez-Sáez et al. (2021) | Hu & Tak (2020) | This paper |
|---|---|---|---|---|
| Motivation | Focus on how well RAE learns the underlying processes and achieves modelling and forecasting of the general trend, rather than the stochasticity. | Aim to recognize changing-look AGN by modelling its light curves with VRAE and obtain their general behaviours and features. | Assuming correlations of $SF_\infty$ and $\tau$ among bands, they proposed a state-space representation of a multivariate DRW model for modelling AGN light curves in five bands and with irregular cadences. | Model the whole 10-yr light curves and predict the vacancies and stochastic features, thus helping LSST cadence strategies evaluation. |
| Data source and input format | Approximately 15 000-decade long quasar light curves from Catalina Real-time Transient Survey; normalized mags and mag_errors, $\Delta t$. | 230 451 various AGNs' light curves from Zwicky Transient Facility data release 5 (ZTF DR5); normalized mags and mag_errors, $\Delta t$. | Generated simulated light curves via two steps: Simulate full light curves using the DRW model, and shape them to realistic sparse light curves. Then they fit them to univariate and multivariate models separately. | Studied ~9000 SDSS Stripe 82 quasars' DRW parameter distributions, which are further referenced for simulating 1200 DRW and DHO quasars in six bands; normalized mags and mag_errors, at each time-step with vacancies (filled with zeros) at some time-steps. |
| Architecture | RAE for forecasting the temporal flux variation of quasar. | VRAE for modelling AGNs; Isolation Forest algorithm as an anomaly detector providing the anomaly score. | Univariate and multivariate DRW models; Maximum likelihood – Kalman Filtering (Kalman 1960) and Bayesian posterior are applied to estimate parameters. | SRNNs for modelling AGN light curves; Univariate Gaussian process regression (GPR) is applied for estimating CARMA parameters. |
| Key results | Applied a RNN Auto-Encoder (AE) to model and predict quasar variability and reported that the AE showed better performance than DRW modelling in forecasting short-term variability. Additionally, they found temporal asymmetry in the optical variability, and the decrease of the amplitude of the variability asymmetry as the luminosity and/or black hole mass increases. | Modelled light curves from the ZTF DR5 with a VRAE architecture obtaining a set of attributes from the VRAE latent space that describes the general behaviour of the data. These attributes are used as features for an anomaly detector. These anomalies are dominated by bogus candidates, identify 75 promising changing-state AGN candidates. | Compared the results of univariate DRW model and multivariate DRW model fittings for simulated light curves and a SDSS spectroscopically-confirmed quasar, indicating that multivariate model can help reveal the possible similarities of true time-scales across five bands. They also estimated the time decay between doubly lensed multiband light curves using the multivariable process. | Applied SRNN to model simulated LSST cadence AGN light curves, learn the stochasticity and latent features, and eventually recover their daily light curves. Provided a metric $M_{err}$ to quantify the influence of different LSST cadence strategies on CARMA parameter recovery. |

### 6.3 Implications in LSST

Our research is expected to help the LSST AGN collaboration to consider optimal cadences for studying AGN variability. To investigate this, we examined the results of our metric evaluation (equation 8) for each LSST band under various classes of proposed LSST cadence (Fig 12). For existed cadence strategies, we conclude that filterdist is the best cadence for band-averaged case. As *u* band always has the worst score due to the fewest total observations, we also only consider a metric with only *g*, *r*, *i*, *z*, and *y* bands, and find that cadence_drive is favoured in this case. If we just look at the 'best' band case, u_long is slightly better than baseline and cadence_drive.

However, our experiments show that SRNN performs better for uniformly-seasonal light curves (see Fig 8) with denser detections, which is reasonable as with more information provided for the SRNN, it can learn the underlying features more easily. Therefore, it might perform better for the DDFs, where the cadences are more dense and uniform compared to the WFD survey.

### 6.4 Model selection and caveats

When LSST data emerges, CARMA models will be fitted to the irregular light curves in Figs 8–10. The fitting procedure for CARMA is MLE (e.g. GPR), or a Bayesian variant (e.g. MCMC) if prior constraints on parameters are available. It is worth noting that in our research, we chose the GPR method as it is much less time-consuming than MCMC. As we have a sufficient amount of data for parameter estimations, the biases caused by GPR can be greatly relieved and the estimation distributions are of our interest.

In standard practice, the choice between AR models, such as DRW - CARMA(1,0) and DHO - CARMA(2,1), should be based on a penalized likelihood measure such as the Akaike Information Criterion (AIC; Akaike 1973). Best-fitting models are calculated for a range of (*p* and *q*) and, using the AIC to achieve model selections. The balance between oversimplicity versus overcomplexity of the models is determined by the data.

Moving away from whether DRW or DHO models are preferred for an observed light curve, goodness-of-fit tests will be needed to show that the best-fitting model is adequate. The Anderson–Darling goodness-of-fit test of the cumulative observed versus model brightness is a reasonable tool (Stephens 1974) as well as more powerful residual diagnostics including the, e.g. Ljung–Box test (Ljung & Box 1978) for autocorrelation and augmented Dickey–Fuller test (Dickey & Fuller 1979) for stationarity. Indeed, time-series analyses will likely move away from the SF (see Emmanoulopoulos, McHardy & Uttley 2010) and likely focus more on the ACF. Ultimately, model selection should allow astrophysical insight into the accretion process (e.g. presence or absence of a harmonic oscillator and non-stationary; Tachibana et al. 2020).

## 6.5 SRNN forecasting

Forecasting future light-curve behaviour for a given object is another goal of applying ML to AGN. We tried to use SRNN for forecasting, and tested its ability to predict one or several steps for one run and fold the outputs as the input for the next run, and to predict a length of future light curves for one run. However, the results are not promising: Our SRNN can only predict light curves for about one month, with low accuracy. As time increases, the error gets bigger and light curves become increasingly flat. This suggests that the SRNN architecture may not be suitable for forecasting, due to its non-deterministic (random) nature. More experiments will be done in the future.

## 7 CONCLUSIONS

Cadence strategy plays an important role in light-curve modelling, especially for AGN with long time-scales and high variability features. This paper introduces an SRNN to model realistic LSST-cadence light curves simulated using CARMA models, and provides a quantitative metric to evaluate and compare the performance between five selected LSST cadence strategies.

In this study, we:

(i) Investigated the popular CARMA models, which are often applied in AGN light-curve simulations. In addition to the usual DRW model, we also explored DHO models (underdamped and overdamped cases), which may be more applicable to some AGN.

(ii) Designed a *QuasarMetric* for simulating LSST-cadence light curves (MJD, mag, and $\sigma_{mag}$) given varying input CARMA parameters.

(iii) Applied modified SRNNs to AGN light-curve modelling, to achieve stochastic modelling and prediction

(iv) Provided a metric, $M_{err}$, for estimating how well SRNN can help recover the input parameters with GPR, compared with pure LSST-cadence light curves.

(v) Concluded that filterdist, cadence_drive, and u_long strategies are the optimal when six bands, five bands, and the best band are considered, respectively.

However, as shown in e.g. Figs 8 and 13, the long gaps inherent in all suggested WFD cadences make it extremely difficult for CARMA models, and the SRNN implementation, to recover accurate variability time-scales. Moreover, in real LSST data most AGN will be fainter than the SDSS sample used to construct our models, and hence photometric noise will introduce further uncertainties. This leads us to conclude that LSST WFD data may not be particularly well suited to AGN variability studies, at least with current methods. Progress may be made by further developing the ML methods, for example by combining our SRNN architecture with the multiband approach of Hu & Tak (2020), or by focussing on the higher cadenced data from the LSST DDFs for a smaller sample of AGN. The research we have presented here provides a method to quantify the performance of any potential cadence strategy, and we expect this will prove useful to the AGN community in preparing for LSST.

## ACKNOWLEDGEMENTS

We are very grateful for the support from Weixiang Yu and Dr. Darren White. MN and XS are supported by the European Research Council (ERC) under the European Union's Horizon 2020 research and innovation programme (grant agreement no. 948381). MN acknowledges a Fellowship from the Alan Turing Institute.

## DATA AVAILABILITY

The code and simulated data underlying this article are available at https://github.com/XinyueSheng2019/LSST_AGN_SRNN_Paper.

# APPENDIX A: THE CARMA(1,0) AND CARMA(2,1) MODELS

## A1 CARMA in a nutshell

Kelly et al. (2014) and Moreno et al. (2019) show the detailed introduction of CARMA in different dimensions. Here, we would like to extract the key information from them, which are beneficial for our research.

CARMA($X$, $Y$) is derived from discrete ARMA model, which is applied to... equation (A1) is the CARMA equation (from equation 1 in Kelly et al. 2014):

$$\frac{d^p y(t)}{dt^p} + \alpha_{p-1}\frac{d^{p-1} y(t)}{dt^{p-1}} + ... + \alpha_0 y(t)$$

$$= \beta_q \frac{d^q \epsilon(t)}{dt^q} + \beta_{q-1}\frac{d^{q-1} \epsilon(t)}{dt^{q-1}} + ... + \epsilon(t). \tag{A1}$$

The left-hand side of the equation with C-AR coefficients $\alpha$ describes the AR part of the system; the right-hand side describes the driving perturbation C-MA. In our work, we only focus on DRW - CARMA(1,0) and DHO - CARMA(2,1). Table A1 shows all relevant equations about DRW and DHO, followed by notation explanation in Table A2.

**Table A1.** Summary of the key DRW and DHO equations (Kelly et al. 2009, 2014; Kasliwal et al. 2017; Moreno et al. 2019). For the RHS of each differential equation, $\epsilon(t)$ corresponds to a white noise process with the mean equal to zero and variance ($\sigma^2_{\mathrm{Noise}}$) equal to one (Kelly et al. 2009). For DRW, there is only one root, $r_1$, which is equal to $-\alpha_1$. For DHO, the $r_1$ and $r_2$ mean the two roots of characteristic equations for the LHS of each differential equation. These could be real or complex, corresponding to overdamped and underdamped cases, respectively. Further symbols are defined in Table A2. It is worth noting that the top four sections define the statistical models while the bottom three sections are non-parametric transforms of the CARMA model.

| CARMA | DRW-CARMA(1,0) | DHO-CARMA(2,1) |
|---|---|---|
| Differential equation | $\mathrm{d}^1 x + \alpha_1 x(t) = \beta_0 \mathrm{d}W(t)$ <br> $\mathrm{d}X(t) = -\frac{1}{\tau}X(t)\mathrm{d}t + \sigma\sqrt{\mathrm{d}t}\,\epsilon(t) + b\mathrm{d}t$ <br> where $\tau$, $\sigma$, and $t > 0$ | $\mathrm{d}^2 x + \alpha_1 \mathrm{d}^1 x + \alpha_2 x = \beta_0(\mathrm{d}W) + \beta_1 \mathrm{d}^1(\mathrm{d}W)$ |
| SSM | $x(t+\delta t) = \mathrm{e}^{\alpha_1 \delta t}x(t) + \beta_0 \int_0^{\delta t} \mathrm{e}^{\alpha_1(\delta t - s)}\mathrm{d}W_s$ | $x(t+\delta t) = \mathrm{e}^{\alpha_1 \delta t}x(t) + \mathrm{e}^{\alpha_2 \delta t}x(t) + \int_0^{\delta t} \mathrm{e}^{A(\delta t - s)}B\mathrm{d}W_s,$ <br> $A = \begin{bmatrix} -\alpha_1 & 1 \\ -\alpha_2 & 0 \end{bmatrix}$, $B = [\beta_1, \beta_0]^T$ |
| Input parameters | $\tau_{\mathrm{decay}} = -\frac{1}{r_1} = \frac{1}{\alpha_1}$ <br><br> $\sigma = \beta_0^2$ | $\tau_{\mathrm{blue}} = \left\| \frac{\beta_1}{\beta_0} \right\|$ <br> $\sigma^2_{\mathrm{lic}} = ACVF(0) = \sigma^2_{\mathrm{Noise}} \frac{\beta_0^2 + \beta_1^2 \alpha_2}{2\alpha_1 \alpha_2}$ <br> Overdamping: <br> $\tau_{\mathrm{rise}} = \frac{1}{r_2 - r_1} ln \left\| \frac{r_1}{r_2} \right\| \approx \left\| \frac{1}{min(r_1, r_2)} \right\|$ <br> $\tau_{\mathrm{decay}} = \left\| \frac{1}{max(r_1, r_2)} \right\|$ <br> Underdamped: <br> $\omega_n = \sqrt{\alpha_2}$ <br> $\omega_d = \omega_n \sqrt{1 - \xi^2} = \sqrt{\alpha_2 - \frac{\alpha_1^2}{4}}$ <br> $\xi = \frac{\alpha_1}{2\sqrt{\alpha_2}}$ <br> $\tau_{\mathrm{QPO}} = \left\| \frac{2\pi}{\omega_n} \right\|$ <br> $T_{\mathrm{dQPO}} = \left\| \frac{2\pi}{\omega_d} \right\|$ <br> $\tau_{\mathrm{decay}} = \left\| \frac{1}{\mathcal{R}(r_1)} \right\| = \left\| \frac{1}{\mathcal{R}(r_2)} \right\| = \left\| \frac{2}{\alpha_1} \right\|$ |
| IR | $\mathrm{d}^1 G + \alpha_1 G = \delta(t)$ <br> $G(t) = C\mathrm{e}^{-\alpha_1 t}$ <br> $\begin{bmatrix} r_1 & r_2 \\ 1 & 1 \end{bmatrix} \begin{bmatrix} C_1 \\ C_2 \end{bmatrix} = \begin{bmatrix} 1 \\ 0 \end{bmatrix}$ | $G + \alpha_1 \mathrm{d}^1 G + \alpha_2 G = \delta(t)$ <br> $G(t) = C_1 \mathrm{e}^{r_1 t} + C_2 \mathrm{e}^{r_2 t}$ |
| ACF | $\mathrm{ACF}(\Delta t) = \mathrm{e}^{-\frac{\|\Delta t\|}{\tau}}$ | $\mathrm{ACVF}(\Delta t) = C_0 \left( C_1 \mathrm{e}^{r_1 \Delta t} + C_2 \mathrm{e}^{r_2 \Delta t} \right)$ <br> $\mathrm{ACF}(\Delta t) = \frac{\mathrm{ACVF}(\Delta t)}{\mathrm{ACVF}(0)}$ <br> $C_0 = \frac{\sigma^2_{\mathrm{Noise}}}{2(r_2 - r_1)}$ <br> $C_1 = -\frac{\beta_0^2 - \beta_1^2 r_1^2}{\mathcal{R}(r_1)(r_1 + r_2^*)}$ <br> $C_2 = \frac{\beta_0^2 - \beta_1^2 r_2^2}{\mathcal{R}(r_2)\left(r_2 + r_1^*\right)}$ |
| SF | $\mathrm{SF}(\Delta t) = \mathrm{SF}_\infty(1 - \mathrm{e}^{-\|\Delta t\|/\tau})^{1/2}$, $\mathrm{SF}_\infty = \sqrt{2}\sigma$ | $\mathrm{SF}(\Delta t) = \sqrt{2(\mathrm{ACVF}(0) - \mathrm{ACVF}(\Delta t))}$ |
| PSD | $p(f) = \frac{\beta_0^2}{\alpha_1 + (2\pi f)^2}$ | $p(f) = \frac{1}{2\pi} \frac{\beta_0^2 + 4\pi^2 \beta_1^2 f^2}{16\pi^4 f^4 + 4\pi^2 \left(\alpha_1^2 - 2\alpha_2\right)f^2 + \alpha_2^2}$ |

**Table A2.** Key parameters in the DRW and DHO models.

**DRW**

| | |
|---|---|
| $\tau$ | The characteristic/decay time-scale. Represents the time for the time series to become uncorrelated. |
| $SF_\infty$ | The long-term variability, $SF_\infty = \sqrt{2}\sigma$. |
| $\sigma$ | The long-term standard deviation of variability, independent of $\tau$ and observation length, however, when the observation length is much longer than $\tau$, the variance of time series will be very close to $\sigma$. |
| $\hat{\sigma}$ | $\hat{\sigma} = SF_\infty/\sqrt{\tau} = \sqrt{2}\sigma/\sqrt{\tau}$ <br> For $\Delta t \ll \tau$ the dispersion between two points is $\hat{\sigma}|\Delta t|^{1/2}$; <br> for $\Delta t \gg \tau$, the dispersion asymptotes to $\sigma$ (Kozłowski et al. 2010). |
| $dW$ & $\epsilon(t)$ | Proved by Ito's theorem (Thomas 1986), Wiener increments $dW$ and random variable $\epsilon(t)$ are equivalent. |

**DHO**

| | |
|---|---|
| $\alpha_1$ | First coefficient of C-AR. |
| $\alpha_2$ | Second coefficient of C-AR, equal to $\omega^2$. |
| $\beta_0$ | First coefficient of MA equation. |
| $\beta_1$ | Second coefficient of MA equation. |
| $\tau_{decay}$ | Decay time-scale for DHO. |
| $\tau_{rise}$ | Rise time-scale for overdamped DHO. It is a feature in Green's function showing the time corresponding to the peak value. When $\tau_{rise} = 0$, it becomes DRW. |
| $\tau_{blue}$ | The ratio of $\left\|\frac{\beta_1}{\beta_0}\right\|$. As $\tau_{blue}$ increases, the time series becomes more erratic. |
| $\omega_n$ | Natural QPO frequency. |
| $\omega_d$ | Decay QPO frequency. |
| $\tau_{QPO}$ | Period of natural QPO frequency. |
| $T_{dQPO}$ | Period of decay QPO frequency. |
| $\xi$ | Damping ratio of the oscillator, equal to $\frac{\alpha_1}{2\sqrt{\alpha_2}}$. <br> $\xi > 1$, it is an overdamped DHO; <br> $\xi = 1$, it is a critical damped DHO; <br> $\xi < 1$ it is an underdamped DHO. |

# APPENDIX B: STOCHASTIC RECURRENT NEURAL NETWORK

## B1 The ELBO

Maximizing $\mathcal{F}$ is the equivalent of maximizing the (log) likelihood, $p(x|z)$, and minimizing the distance to the prior, $D_{KL}(q(z)\|p(z|x))$.

KL divergence is always equal to, or greater than 0. Hence the ELBO is always below the (log) evidence of our data and in that sense it is the lower bound: $\mathcal{F}(\theta, \phi) = \sum_i \mathcal{F}_i(\theta, \phi) \leq \mathcal{L}(\theta)$.

The KL divergence is for measuring the differences between two probability distributions. In this research, it is applied to measure the differences between prior and posterior $z$ distributions. For one pair of input and output sequences, the calculation is shown in equation (B1).

$$D_{KL}(q(z) \| p(z|x)) = \int_z q(z)\log \frac{q(z)}{p(z|x)}dz$$

$$= \mathbb{E}_{z\sim q}[\log q(z)] - \mathbb{E}_{z\sim q}[\log p(z|x)]. \tag{B1}$$

## B2 SRNN implementation

There is no simple command to invoke the SRNN in TensorFlow. Instead, as described in Section 4.1, the SRNN is built with a generative model and an inference model. This is implemented for our study as follows:

```
########## Generative Model ###########
## two layers
X_inputs = Input(shape = (timestep, features),
  name = 'X_input')
h1_out = Bidirectional(GRU(num_neutron, name =
  'h_0', return_sequences=True))(X_inputs)
h1_out = Bidirectional(GRU(num_neutron, name =
```

```
  'h_0', return_sequences=True))(d1_out)

## Bidirectional wrapper for RNN;
## generate prior z from h1
prior_z_mean = Bidirectional(GRU(z_neutron, name
  ='prior_mean', return_sequences=True))
  (h1_out)
prior_z_log_var = Bidirectional(GRU(z_neutron,
  name = 'prior_log_var',
  return_sequences=True))(h1_out)
prior_sampled_z = Sampling(name =
  'sampling_prior')([prior_z_mean,
  prior_z_log_var])
merged_prior = concatenate([h1_out,
  prior_sampled_z])

## prior outputs will be the real outputs for
  validation set
prior_outputs = TimeDistributed(Dense(1),
  name='time_dist')(merged_prior)

generative_model = Model(X_inputs,
  [prior_outputs, prior_z_mean, prior_z_log_var,
  h1_out], name = 'generative_model')

########## Inference Model ###########
## reverse layer a for inference
g1_out = Input(shape = (pred_timestep,
  num_neutron2), name = 'g1_out')
Y_inputs = Input(shape = (pred_timestep,
  pred_features), name = 'Y_input')
merged_input = concatenate([g1_out, Y_inputs])
```

```
# Reversed Layer, a
a_out = Bidirectional(GRU(num_neutron,
  name = 'a_0', return_sequences=True,
  go_backwards=True))(merged_input)

posterior_z_mean = Bidirectional
  (GRU(z_neutron, name = 'posterior_z_mean',
  return_sequences=True, go_backwards=True))
  (a_out)
posterior_z_log_var = Bidirectional
  (GRU(z_neutron, name = 'posterior_z_log_var',
```

```
  return_sequences=True, go_backwards=True))
  (a_out)
inference_model = Model([g1_out, Y_inputs],
  [posterior_z_mean, posterior_z_log_var],
  name = 'inference_model')

## Fulfiling the SRNN
srnn = SRNN(generative_model, inference_model)
```

This paper has been typeset from a TeX/LaTeX file prepared by the author.